\documentclass[a4paper,12pt]{article}
\usepackage[a4paper,text={16.8cm,22.4cm}]{geometry}

\usepackage{amsmath,amssymb,bm}
\usepackage{psfrag}
\usepackage{graphicx, graphics}

\newcommand{\scetopi}{\text{1PI}_{\text{SCET}}}
\newcommand{\scetpim}{\text{PIM}_{\text{SCET}}}
\newcommand{\msbar}{\overline{\text{MS}}}
\def\lb#1{\if 1#1 \ln\beta \else \ln^#1\beta \fi}
\usepackage{color}

\def\beq{\begin{equation}}
\def\eeq{\end{equation}}
\def\beqa{\begin{eqnarray}}
\def\eeqa{\end{eqnarray}}

\begin{document}

%\title{Insert your title here}
%%\subtitle{Do you have a subtitle?\\ If so, write it here}
%\author{First author\inst{1} \and Second author\inst{2}% etc
%% \thanks is optional - remove next line if not needed
%%\thanks{\emph{Present address:} Insert the address here if needed}%
%}                     % Do not remove
%%
%\offprints{}          % Insert a name or remove this line
%%
%\institute{Insert the first address here \and the second here}
%%
%\date{Received: date / Revised version: date}
%% The correct dates will be entered by Springer
%%
%\abstract{
%Insert your abstract here.
%%
%\PACS{
%      {PACS-key}{discribing text of that key}   \and
%      {PACS-key}{discribing text of that key}
%     } % end of PACS codes
%} %end of abstract
%%
%\maketitle

\begin{titlepage}

  \begin{flushright}
   \today \\
  \end{flushright}

  \vspace{5ex}
  
  \begin{center}
    \textbf{Top-quark production and QCD} \vspace{7ex}
    
    \textsc{Nikolaos Kidonakis$^a$ and 
      Ben D. Pecjak$^b$} \vspace{2ex}
  
    \textsl{${}^a$ Kennesaw State University, Physics \#1202 \\
      Kennesaw, GA 30144, USA\\[0.3cm]
     ${}^b$Institut f\"ur Physik (THEP), Johannes Gutenberg-Universit\"at\\
      D-55099 Mainz, Germany}
  \end{center}

  \vspace{4ex}

\begin{abstract}
   
We review theoretical calculations for top-quark production that include 
complete next-to-leading-order QCD corrections as well as higher-order 
soft-gluon corrections from threshold resummation. We discuss in detail the 
differences between various approaches that have appeared in the literature 
and review results for top-quark total cross sections and differential 
distributions at the Tevatron and the LHC.

\end{abstract}

\end{titlepage}

\section{Introduction} 

Top-quark physics is a centerpiece of the research programs at the Tevatron
and the LHC.  Due to its unique position as the heaviest particle known 
to date, the top quark not only plays a role in many models of physics 
beyond the Standard Model (SM), but also in the precision electroweak
fits constraining the mass of the Higgs boson.

The dominant mechanism for creating top quarks at hadron colliders
is the production of a top-antitop pair through QCD interactions.
Tens of thousands of top-quark pairs have already been produced and studied
at the Tevatron since the discovery of the top quark there in 1995 
\cite{Abe:1995hr,Abachi:1995iq}, and millions will be produced at the 
LHC.  As the measurements become more precise, it will become increasingly
important to have reliable QCD predictions of the total and differential
pair-production cross sections.  This is true not only if 
signals of new physics manifest as slight excesses in differential cross
sections, but also if top-quark properties are completely
determined through the SM. In that case top-quark pair production
will be a benchmark process,  used in tasks from subtracting 
backgrounds to constraining gluon PDFs in regions of $x$ relevant
for Higgs production.

Another process, with a smaller cross section than for top-antitop, is
single top-quark production, which can proceed via three distinct partonic 
channels. Single top-quark production is important in probing electroweak 
theory and studying the electroweak properties of the top quark as well as 
for the discovery of new physics, since the top-quark mass is of the same 
order of magnitude as the electroweak symmetry breaking scale.

The main purpose of this review is to summarize certain aspects of QCD
calculations for the total and differential pair-production cross sections. We
will focus on inclusive observables $pp(\bar p)\to t \bar t X$, summed over
spins, although some results for the single top-quark total cross section in
the $t$-channel, $s$-channel, and in $tW$ production are also discussed.  In
particular, we will perform detailed discussions of the total inclusive
production cross section, the top-pair invariant mass distribution, the
transverse momentum ($p_T$) distribution of the top (or antitop) quark, and
the rapidity distribution, as well as the forward-backward (FB) asymmetry at
the Tevatron.  While this leaves out a number of interesting topics related to
other exclusive observables, these are covered in excellent reviews of
top-quark physics in the literature, e.g. \cite{Bernreuther:2008ju}.  In
this review we present in detail the most up-to-date theoretical predictions
for the observables mentioned above.

The starting point for a study of inclusive observables in top-quark
pair production is the next-to-leading-order (NLO) calculation of
differential and total cross sections performed more the two decades
ago \cite{Nason:1987xz,Nason:1989zy,Beenakker:1988bq,Beenakker:1990maa}. 
However, to keep up with experimental precision
requires to go beyond them.  Many theorists have responded to this
challenge by tackling parts of the diagrammatic calculations needed to
reach next-to-next-to-leading-order (NNLO) accuracy and a large number of papers 
have appeared with pieces of the NNLO calculation.  Due to very recent
progress it now looks feasible that the NNLO cross section may be
available in the near future, a step forward that will mark a major
accomplishment.

In the absence of the full NNLO results, an important tool for including
higher-order corrections is soft gluon resummation at next-to-leading-logarithm (NLL) accuracy \cite{Kidonakis:1996aq,Kidonakis:1997gm} and beyond. 
The resummation techniques apply to both differential and total 
cross sections and fixed-order expansions were derived from the 
NLL resummed cross section with additional subleading logarithms in  
\cite{Kidonakis:2003qe,Kidonakis:2008mu}. 
 Recently,  next-to-next-to-leading-logarithm (NNLL) predictions
have appeared for the total cross section \cite{Langenfeld:2009wd,Ahrens:2010zv,Kidonakis:2010dk,Ahrens:2011mw} and for double differential 
cross sections with respect to the top quark $p_T$ and rapidity \cite{Kidonakis:2010dk,Ahrens:2011mw,Kidonakis:2011zn}, 
or the pair invariant mass and rapidity \cite{Ahrens:2010zv}.
A primary goal of this review is to clarify the 
similarities and differences between the different approaches.

The remainder of this document is organized as follows.  
In Section~\ref{sec:fixedorder} we review the basics
of fixed-order calculations to NLO and summarize 
progress up to NNLO.  In Section~\ref{sec:resummation},
we discuss the underlying assumptions and techniques
in soft gluon resummation, providing a theory comparison 
of different approaches.  We present numerical
results for the  total inclusive cross section 
within various approaches in Section~\ref{sec:totalcs},
for a selected group of differential cross sections
in Section~\ref{sec:diffcs}, and for the FB asymmetry
in Section~\ref{sec:FBasy}.  Single top production 
is briefly discussed in Section~\ref{sec:singletop},
and some closing remarks are made in Section~\ref{sec:conclusions}.

\section{Kinematics and fixed-order results}
\label{sec:fixedorder} 
In this section we discuss the basics of fixed-order calculations 
for inclusive top-quark pair production at hadron colliders.  
We thus consider the scattering process
\begin{align}
  \label{eq:process}
  N_1(P_1) + N_2(P_2) \to t(p_3) + \bar{t}(p_4) + X \, ,
\end{align}
where $N_1$ and $N_2$ indicate the incoming protons (LHC) or proton and
anti-proton (Tevatron), and $X$ represents an inclusive hadronic final
state.  The calculation of the differential cross section
relies  on factorization in QCD.  This is the statement that a generic
differential hadronic cross section is given in terms of 
partonic cross sections associated with scattering of gluons
and quarks, after a convolution integral with parton distribution
functions (PDFs) that describe the parton content of the hadrons.  
We write this schematically as 
\begin{align}
\label{eq:factorization}
d\sigma = \sum_{ij\in{q,\bar q,g}} \int dx_1 dx_2 \, \phi_{i/N_1}(x_1,\mu_F) \, 
\phi_{j/N_2}(x_2,\mu_F) \, 
d\hat{\sigma}_{ij}(x_1,x_2,\mu_F,\mu_R,\alpha_s(\mu_R)),
\end{align}
where the $x_i$ are longitudinal momentum fractions of the incoming
partons $i$, $\phi_{i/N}$ are PDFs, and the $d\hat{\sigma}_{ij}$ are
partonic cross sections.  The collinear singularities are factorized
in a process-independent manner and absorbed into the PDFs, which
depend on the factorization scale $\mu_F$.  
The partonic cross sections  can be expanded in a fixed-order 
series in the strong coupling constant $\alpha_s(\mu_R)$ as 
\begin{align}
d\hat{\sigma}_{ij} = \alpha_s^2\left[d \hat{\sigma}_{ij}^{(0)} +
 \frac{\alpha_s}{\pi} d \hat{\sigma}_{ij}^{(1)} + 
 \frac{\alpha_s^2}{\pi^2} d \hat{\sigma}_{ij}^{(2)}
\dots      \right] \, ,
\end{align}
where the first term in square brackets is referred to as
leading-order (LO), the second term next-to-leading-order (NLO), the
third term next-to-next-to-leading-order (NNLO), and so on.  The
physical cross section is independent of the
factorization scale $\mu_F$ and the renormalization scale $\mu_R$.
However, the truncation of the infinite perturbative series at finite
order typically results in a non-negligible numerical dependence.

\begin{figure}%[h!]
\begin{center}
\includegraphics[width=.2\textwidth]{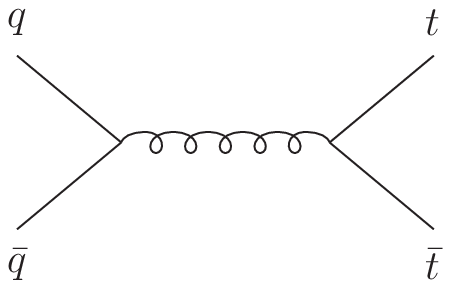}   
\hspace*{15mm}  
\includegraphics[width=.68\textwidth]{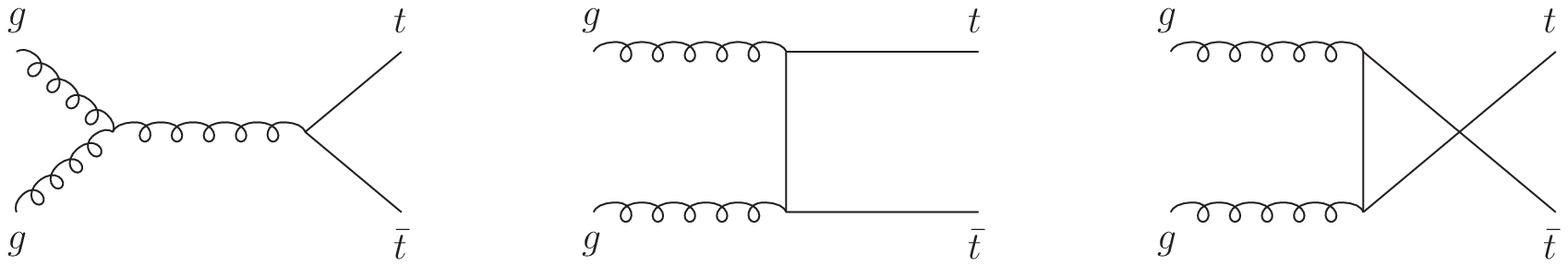}
\end{center}
\vspace{-2mm}
\caption{ The four Feynman diagrams contributing
to top-quark pair production at LO.\label{fig:born} }
\end{figure}

 At LO, the partonic cross sections receive contributions from the
quark-antiquark annihilation and gluon fusion subprocesses 
\begin{align}
  \label{eq:partprocess}
  q(p_1) + \bar{q}(p_2) &\to t(p_3) + \bar{t}(p_4) \, , \nonumber
  \\
  g(p_1) + g(p_2) &\to t(p_3) + \bar{t}(p_4)  \, .
\end{align}
The Feynman diagrams needed for the computation of the differential
cross section are shown in  Figure~\ref{fig:born}.

%%%%%%%%%%%%%%%%%%%%%%%%%%%%%%%%%%%%%%%%%%%%%%%%%
%%%%%%%%%%%%%%%%%%%%%%%%%%%%%%%%%%%%%%%%%%%%%%%%%

The results for the LO differential partonic cross sections read
\begin{align}
\label{eq:born}
\hat{s}^2 \, \frac{d\hat\sigma_{q\bar q}^{(0)}}{d\hat{t}_1 d\hat{u}_1} &= 
\frac{\pi C_F}{N_c} \left( \frac{\hat{t}_1^2+\hat{u}_1^2}{\hat{s}^2}
    + \frac{2m_t^2}{\hat{s}} \right)\delta(\hat{s}+\hat{t}_1+\hat{u}_1) , 
\nonumber \\
\hat{s}^2\, \frac{d\hat\sigma_{g g}^{(0)}}{d\hat{t}_1 d\hat{u}_1} &= 
\frac{2\pi N_c C_F}{(N_c^2-1)^2}
\left( C_F \frac{\hat{s}^2}{\hat{t}_1\hat{u}_1} - C_A
  \right) \left[ \frac{\hat{t}_1^2+\hat{u}_1^2}{\hat{s}^2} + \frac{4m_t^2}{\hat{s}} 
- \frac{4m_t^4}{\hat{t}_1\hat{u}_1}
  \right]\delta(\hat{s}+\hat{t}_1+\hat{u}_1) ,
\end{align}
where $C_F=(N_c^2-1)/2N_c$ and $C_A=N_c$, with $N_c=3$ the number of colors
in QCD.  The partonic Mandelstam variables are defined as
\begin{align}
\label{eq:mandelstam}
 \quad \hat{s}=(p_1+p_2)^2 \, , \quad \hat{t}_1=(p_1-p_3)^2-m_t^2 \, ,
  \quad \hat{u}_1=(p_2-p_3)^2 -m_t^2 \, ,
\end{align}
where $m_t$ is the top quark mass.
The  partonic momenta are related to the hadronic ones through
$p_1=x_1P_1$ and $p_2=x_2P_2$, so the connection between the
hadronic invariants
\begin{align}
  \label{eq:hadmandelstam}
  s &= (P_1+P_2)^2 \, , \quad t_1 = (P_1-p_3)^2-m_t^2 \, , 
\quad u_1=(P_2-p_3)^2 -m_t^2 \, ,
\end{align}
and the partonic ones is
\begin{gather}
  \hat{s} = x_1 x_2 s \, , \quad \hat{t}_1 = x_1 t_1 \, , 
\quad \hat{u}_1= x_2 u_1 \, .
\label{eq:mandelstampart}
\end{gather}

At Born level, any differential cross section can be derived from
(\ref{eq:born}) by an appropriate change of variables. 
For instance, single-particle-inclusive (1PI) cross
sections depending on the rapidity $y$ and $p_T$ of the top quark in the
laboratory frame are calculated using
\begin{align}
 t_1 = -\sqrt{s} \,  m_\perp\, e^{-y} \, , 
\quad u_1 = -\sqrt{s} \, m_\perp e^{y} \, ,
\end{align}
where $m_\perp = \sqrt{p_T^2 + m_t^2}$,
while pair-invariant-mass (PIM) observables depending on 
the pair-invariant mass $M_{t\bar t}^2\equiv (p_3+p_4)^2$ and
scattering angle of the top-quark are obtained using
\begin{align}
  \label{eq:tu}
\hat{s}=M_{t\bar t}^2\, , \qquad   \hat{t}_1 = -\frac{M_{t\bar t}^2}{2} ( 1 - \beta_t
  \cos\theta ) \, , \qquad \hat{u}_1 = -\frac{M_{t\bar t}^2}{2} ( 1 +
  \beta_t \cos\theta ) \, .
\end{align}
where 
$\beta_t = \sqrt{1-4m_t^2/M_{t\bar t}^2}$.  

The total cross cross section itself
is conventionally written in the form 
\begin{align}
\label{eq:ff}
  \sigma(s,m_t) = \frac{\alpha_s^2(\mu_R)}{m_t^2} \sum_{i,j} \int_{4m_t^2}^s \frac{d\hat{s}}{s}
  \, \Phi_{ij} \bigg( \frac{\hat{s}}{s}, \mu_F \bigg) \, f_{ij} \bigg( \frac{4
    m_t^2}{\hat{s}}, \mu_F, \mu_R \bigg) \, ,
\end{align}
where  the parton  luminosities are defined as
\begin{align}
  \Phi_{ij}(y,\mu_F) = \int_y^1 \frac{dx}{x} \, 
\phi_{i/N_1}(x,\mu_F) \, \phi_{j/N_2}(y/x,\mu_F) \,
  \, .
\end{align}
The perturbative scaling functions $f_{ij}$ are easily obtained by integrating
the differential partonic cross sections over the appropriate phase
space.  We define their fixed-order expansion as
\begin{align}
  \label{eq:fexp}
  f_{ij} &= f_{ij}^{(0)} + 4\pi \alpha_s f_{ij}^{(1)} 
+ \left(4\pi\alpha_s\right)^2 f_{ij}^{(2)} + \dots .
\end{align}
These scaling functions are dimensionless quantities normally written 
in terms of
\begin{align}
\label{eq:betadef}
\beta = \sqrt{1-\frac{4m_t^2}{\hat{s}}}, \qquad \rho \equiv 1-\beta^2,
\end{align}  
where at Born level $\beta$ is just the magnitude of the three-velocity of the
top or antitop quark.
For reference, the LO expressions in the two channels read
\begin{align}
\label{eq:fijborn}
f_{q\bar q}^{(0)}&= \frac{\pi C_F\beta \rho}{12 N_c}\left[2+\rho\right]  \\
f_{g g}^{(0)} & = \frac{\pi N_c C_F \beta \rho}{(N_c^2-1)^2}
\left\{\frac{1}{\beta}\left[C_F\left(1+\rho-\frac{\rho^2}{2}\right)
+C_A\frac{\rho^2}{4}\right]\ln\left(\frac{1+\beta}{1-\beta}\right)
-C_F(1+\rho) -\frac{C_A}{12}(4+5\rho)\right\}\, .
\end{align}

Predictions made using the LO partonic cross sections depend very strongly on
the choice of the factorization and renormalization scales and are not
appropriate for any sort of detailed phenomenology.  However, they are
sufficient for understanding most of the qualitative features of the
observables discussed later on.  For instance, the LO results predict a total
cross section of roughly $7$~pb at the Tevatron, with a dominant contribution
of around 90\% from the $q\bar q$ channel.  At the LHC with $\sqrt{s}=7$~TeV
the cross section is around twenty times as large, and
the  dominant contribution, of around 75\%, is  from the $gg$ channel.  
Obviously, the size of the cross section and relative strength of the
different channels reflects the differences in the parton luminosities at the
two colliders, since the partonic cross sections are the same.  

To make quantitative comparisons between theory and experiment
requires the partonic cross sections at NLO accuracy.  The total and
differential inclusive production cross sections were calculated at
NLO more than 20 years ago
\cite{Nason:1987xz,Nason:1989zy,Beenakker:1988bq,Beenakker:1990maa}.
All of the NLO results for top-quark pair production are implemented
in parton Monte Carlo programs such as MCFM \cite{Campbell:2000bg} or
the NLO version of MadGraph/MadEvent based on \cite{Alwall:2007st,
  Frederix:2009yq}, or also including parton showers in programs such
as MC@NLO \cite{Frixione:2003ei}.  These programs allow for the
calculation of arbitrary differential distributions and are thus an
important tool for experimentalists and theorists involved in
phenomenological applications.

Numerical results using the partonic cross sections at NLO will be
reviewed in the following sections. It is conventional to estimate the
theoretical uncertainties associated with the truncation of the
perturbative series to NLO by picking default values of the
factorization and renormalization scales, varying them up and down
by factors of two, and taking the resulting spread of values for the
cross section as an indication of the size of uncalculated, higher
order terms.  Although significantly more stable than LO predictions
under such scale variations, NLO calculations still suffer from scale
uncertainties of roughly 10-20\%.  Higher-order calculations using
soft-gluon resummation reduce the scale uncertainty and will be
covered in the next section.  An even more ambitious approach is to
calculate the full NNLO corrections, and we now review recent progress
in this area.

The calculation of the NNLO cross sections involves two sets of
corrections: \textit{i) virtual corrections}, which can be split into
genuine two-loop or one-loop squared contributions, and \textit{ii)
  real radiation}, which involves either one-loop diagrams with the
emission of one extra parton in the final state, or tree-level
diagrams with two extra partons in the final state.  Needless to say,
the NNLO calculation is very challenging, and so far only pieces of it
are available.

For the virtual corrections, the current situation can be summarized as
follows. Analytic results for the two-loop contributions in the high-energy
limit are known from \cite{Czakon:2007ej, Czakon:2007wk}, along with
contributions to the $q\bar q$ channel from fermionic \cite{Bonciani:2008az}
and planar corrections \cite{Bonciani:2009nb}, and to the $gg$ channel from
planar corrections \cite{Bonciani:2010mn}.  The exact results for the two-loop
contributions in the $q\bar q$ channel are known numerically from
\cite{Czakon:2008zk}.  In addition, the two-loop infrared singularity
structure in the $q\bar q$ and $gg$ channels was determined in
\cite{Ferroglia:2009ep, Ferroglia:2009ii}.  Finally, one-loop squared terms
were calculated in \cite{Korner:2008bn, Anastasiou:2008vd, Kniehl:2008fd}. 

For the real corrections, the pieces arising from one-loop virtual-real graphs
are known from calculations of $t\bar t$ production in association with jets
\cite{Dittmaier:2007wz, Dittmaier:2008yq, Bevilacqua:2010ve, Melnikov:2010iu}.
The more involved task of developing a new subtraction scheme and calculating
the contributions from double real radiation has been dealt with in
\cite{Czakon:2010td, Czakon:2011ve, Anastasiou:2010pw, Abelof:2011jv,
Bernreuther:2011jt}. In light of these advances, the completion of the NNLO
calculation now looks feasible, although a significant amount of work is still
required to assemble all the elements.  In the meantime, approximations of the
NNLO corrections are available using techniques from soft gluon resummation.

We end this section with a couple of comments on progress in NLO
calculations which is slightly outside the main stream of presentation
but nonetheless very important.  So far, we have considered only
inclusive top-pair production with a final state $t\bar t X$. In
reality, the top quark decays almost as soon as it is produced, and
experiment observes the decay products rather than the top quarks
themselves.  If the theory calculations stop at the level of on-shell
top-quark production, experimentalists must fill in the gaps by
correcting their measurements of the decay products (with a possibly
complicated set of acceptances) back to that level.  Obviously, this
should be done as accurately as possible.  Recent calculations of
top-quark production at NLO, including the top-quark decay in the
dominant $t \to b W$ channel (along with leptonic decay of the $W$),
allow for a much more direct comparison between theory and experiment.
The most advanced computations \cite{Denner:2010jp, Bevilacqua:2010qb}
take into account the full correlations between production and decay,
without the further assumption of on-shell top-quark production used
in the narrow-width approximation.  A main result of those works is
that the narrow-width approximation is valid to the percent level for
sufficiently inclusive observables (including those studied in this
review). This means that in most cases one may use equally well the
previous calculations of \cite{Bernreuther:2004jv, Melnikov:2009dn},
which include a complete description of production and decay to NLO
within the narrow-width approximation.

\section{Soft gluon resummation}
\label{sec:resummation}
So far we have considered only the fixed-order perturbative expansion for the
partonic cross sections.  However, there are certain cases where such an
expansion is not the optimal calculational procedure, due the presence of large
logarithms related to soft gluon emission.  The reorganization of the
perturbative expansion in these cases is carried out through techniques referred
to as soft gluon resummation.  

Soft gluon resummation was first applied to top-quark production 
at leading-logarithm (LL) accuracy twenty years ago in 
\cite{Laenen:1991af}. 
At that level the resummation 
is universal and only depends on the identity of the incoming partons 
(quarks and gluons) and does not depend on the details of the hard 
scattering. 
Other approaches at LL followed in 
\cite{Berger:1995xz,Berger:1996ad} and 
\cite{Catani:1996dj}.
However, beyond LL the color structure 
of the process directly enters the resummation. 
Resummation for top quark production at NLL was first presented  
in \cite{Kidonakis:1996aq,Kidonakis:1997gm} where the one-loop soft 
anomalous dimension matrices were calculated. 
NNLO expansions of the differential resummed cross section,  incorporating
additional subleading logarithms, were presented in 1PI and PIM kinematics
in \cite{Kidonakis:2003qe}.
A different approach at NLL for the total cross section only, 
using production threshold, appeared 
in \cite{Bonciani:1998vc}, and was used in \cite{Cacciari:2008zb} 
and with additional subleading logarithms (and Coulomb terms) 
in \cite{Moch:2008qy}. 

Further work on two-loop calculations for soft and collinear singularities
in massless gauge theories \cite{Aybat:2006wq,Aybat:2006mz,Dixon:2008gr,Becher:2009cu,Becher:2009qa,Gardi:2009qi}
and with massive quarks \cite{Kidonakis:2009ev,Mitov:2009sv,Becher:2009kw,Beneke:2009rj,Czakon:2009zw,Ferroglia:2009ep,Ferroglia:2009ii,Ahrens:2010zv,Mitov:2010xw,Beneke:2010da,Kidonakis:2010dk,Ahrens:2011mw} 
have eventually allowed NNLL resummation for $t{\bar t}$ production  
\cite{Langenfeld:2009wd,Ahrens:2010zv,Kidonakis:2010dk,Ahrens:2011mw,Kidonakis:2011zn}.

A typical example where resummation is needed is the top-pair invariant
mass distribution at large values of the invariant mass $M_{t\bar t}$.  To
understand why this is the case, consider the factorized expression for the
hadronic cross section, which reads
\begin{align}
\label{eq:dsigmadmtt}
  \frac{d\sigma}{dM_{t\bar t}} &= 
\sum_{i,j} \int_{\tau}^1
  \frac{dz}{z} \, \Phi_{ij}(\tau/z,\mu_F) \, 
\frac{d\hat\sigma_{ij}(z,M_{t\bar t},\mu_F, \mu_R, \alpha_s(\mu_R))}{dM_{t\bar
      t}} \,.
\end{align}
In the limit of very large invariant mass, the variable $\tau \equiv M_{t\bar
t}^2/s \to 1$, which implies that the partonic variable $z \equiv M_{t\bar
t}^2/\hat{s} \to 1$. In that limit, the partonic center-of-mass energy is just
above the threshold required to produce the top pair with a given invariant
mass, which forces any additional radiated partons to be soft.  Such soft
radiation is associated with IR divergences in the QCD scattering
amplitudes, and after collinear subtractions and cancellations 
with singularities from the virtual corrections, the cross section
contains logarithmic plus-distribution corrections. 
For the $n$-th  order correction to the partonic cross sections in the 
$gg$ and $q\bar q$ channels, these are of the form
\begin{align}
\label{eq:logcors}
 \alpha_s^n \left[ \frac{\ln^m(1-z)}{1-z} \right]_+ ; 
\quad m = 0,\cdots,2n-1 
\, ,
\end{align}
where the plus distributions are defined as 
\begin{align}
\label{eq:plusz}
  \int_\tau^1 dz \left[ \frac{\ln^m(1-z)}{1-z} \right]_+ g(z) = \int_\tau^1 dz \,
  \frac{\ln^m(1-z)}{1-z} \, [g(z)-g(1)]+\frac{g(1)}{m+1}\ln^{m+1}(1-\tau)
\end{align}
for an arbitrary function $g(z)$.  The $qg$ channel also receives up
to double logarithmic corrections but it is suppressed by a relative
factor of $(1-z)$ compared to the $gg$ and $q\bar q$ channels.  

The large logarithms appearing in the limit $z\to 1$ make the
perturbative series poorly behaved, and to make reliable predictions
requires that they be resummed. Because the logarithms are related to
the fact that at the partonic threshold real gluon emission is soft,
such a resummation is interchangeably referred to as ``soft gluon'' or
``threshold'' resummation. The technical machinery required to perform
soft gluon resummation will be reviewed in following two subsections.
For now, we just note that the plus-distribution corrections stem from
terms depending on the dimensionless ratio $2E_s/\mu$, with
$2E_s=\hat{s}(1-z)$ the energy of the soft-gluon radiation in the
partonic center-of-mass frame, so the renormalization-group is the
basic tool for resummation.

While interesting from the technical point of view, the situation
described above is of little phenomenological importance. The reason
is that the differential cross section is essentially zero in the
limit $\tau \to 1$, since in that case the combined energy of the top-quark
pair approaches the collider energy and the probability that two
initial state partons carry such a large fraction of the energy is
tiny.  Both at the Tevatron and at the LHC, the invariant mass
distribution is largest in regions of phase space where $\tau$ is
closer to zero than to one, and the partonic cross section in the
convolution integral (\ref{eq:dsigmadmtt}) is evaluated at values of
$z$ far from unity.  In that case, the leading terms in the soft limit
give the largest contributions to the hadronic cross section if the
parton luminosities as a function of $\tau/z$ are by far largest at
$z\to 1$. If that is true, then the expansion of the partonic cross
section in the $z\to 1$ limit under the integral (\ref{eq:dsigmadmtt})
is parametrically justified.  To a good degree this is an actual
property of the parton luminosities, and detailed studies show that,
at least at NLO, the leading terms in the $z\to 1$ limit provide an
excellent approximation to the full result. Assuming that also beyond
NLO the leading terms in the soft limit account for the bulk of the
corrections, then predictions using soft gluon resummation are an
improvement on the fixed-order expansion even in regions of phase
space where $\tau$ is not close to unity.

The line of reasoning used for the invariant mass distribution can be
applied to soft gluon resummation for other differential partonic
cross sections, or else to the total partonic cross section directly.
In fact, the vast literature on soft gluon resummation in top-quark
pair production can be broken down into the three main cases shown in
Table~\ref{tab:softlimits}.  In each case, one considers the soft
limit of the (differential) partonic cross section shown in the table,
resums corrections which become large in that limit, and relies on the
fall-off of the parton distributions away from the soft limit to
dynamically enhance the partonic threshold region.  The cases of PIM
and 1PI kinematics were first considered in 
\cite{Kidonakis:1996aq,Kidonakis:1997gm} and \cite{Laenen:1998qw}, 
respectively, and work at the level of double differential cross sections.  The
logarithmic corrections in PIM kinematics are of the form
(\ref{eq:plusz}), while in 1PI kinematics they are of form \beq
\alpha_s^n \left[\frac{\ln^m(s_4/m_t^2)}{s_4}\right]_+ \,\quad m =
0,\cdots,2n-1 , \eeq where $s_4={\hat s}+{\hat t}_1+{\hat u}_1$ and the plus
distributions are defined as
\begin{align}
 \int_0^{s_4^{\text{max}}} \left[ \frac{1}{s_4} \ln^n \left(\frac{s_4}{m_t^2}\right) 
  \right]_+ g(s_4) = \int_0^{s_4^{\text{max}}} ds_4 \, \frac{1}{s_4} \ln^n\!\bigg( 
    \frac{s_4}{m_t^2} \bigg) \left[ g(s_4) - g(0) \right] + \frac{g(0)}{n+1} \ln^{n+1}
  \!\bigg( \frac{s_4^{\text{max}}}{m_t^2} \bigg) \,.
\label{1piplus}
\end{align}
It is very important to emphasize that resummed results in 1PI and PIM 
kinematics are obtained after integrating over a specific portion of 
the fully differential phase space with respect to additional soft gluon
radiation.  This means that the results in 1PI or PIM kinematics
apply {\it only} to the differential cross sections in the table, in other
words they are not related by a simple change of variables.  However, 
they can both be used to predict the total cross section by integrating
over the distributions.  Alternatively, one can perform resummation for
the total partonic cross section directly working in the production 
threshold limit, which is the third entry in the table.  In that 
case the large corrections in the soft limit involve simple logarithms
of $\beta$ instead of singular plus distributions. 

As far as soft gluon resummation is concerned, results in the
production threshold limit are actually a special case of PIM and 1PI
kinematics and thus do not contain independent information.  For
instance, the limit $(1-z)\to 0$ plus the additional phase-space
restriction $M_{t\bar t} \to 2m_t$, or $s_4 \to 0$ plus the
restriction $\hat{u}_1+\hat{t}_1 = -4m_t^2$, both imply the limit
$\beta \to 0$.  However, in the limit $\beta \to 0$ there are
additional Coulomb singularities of the form $\ln^m\beta/\beta^n$,
not all of which are determined by soft gluon resummation alone.  In
comparing results for the total cross section obtained in the
different limits, an important consideration is whether the subleading
terms in $\beta$ contained in the 1PI and PIM results, or the Coulomb
singularities are more important.  This is a numerical question which
we will come back to later on.  In the meantime, we pause to explain
in more detail the technical formalism needed to perform resummation
within the different types of kinematics.

\begin{table}
\begin{center}
  \begin{tabular}{c|c|c}
    Name & Observable & Soft limit \\ \hline 
pair-invariant-mass (PIM) &
    $d\sigma/dM_{t\bar t}d\theta$ & $(1-z) = 1-M_{t\bar t}^2/\hat{s}\to 
    0$ \\ 
single-particle-inclusive (1PI) & $d\sigma/dp_T dy$ & $s_4
    = \hat{s}+\hat{t}_1+\hat{u}_1 \to 0$ \\ 
production
    threshold & $\sigma$ & $\beta = \sqrt{1-4m_t^2/\hat{s}}\to 0$ \\
\hline
\end{tabular}
\end{center}
\vspace{-2mm}
\caption{\label{tab:softlimits} 
  The three cases in which soft gluon resummation has been applied.
  The first column indicates the name often used in the literature,
  the second the observable to which it applies, and the third
  the partonic variable associated with large logarithmic 
  corrections in the soft limit.}
\end{table}

 \subsection{Mellin-space resummation}
\label{sec:mellin}

The resummation of the threshold logarithms has traditionally been performed 
in Mellin moment space.
By taking moments, divergent distributions in $1-z$ (PIM kinematics) or 
$s_4$ (1PI kinematics) produce 
powers of $\ln N$, with $N$ the moment variable:
\beq
\int_0^1 dz\; z^{N-1}\left[{\ln^m(1-z)\over 1-z}\right]_+
={(-1)^{m+1}\over m+1}\ln^{m+1}N +{\cal O}\left(\ln^{m-1}N\right)\, .
\eeq
We define moments of the partonic cross section by
${\hat\sigma}(N)=\int dz \, z^{N-1} {\hat\sigma}(z)$ (PIM) or by  
${\hat\sigma}(N)=\int (ds_4/{\hat s}) \;  e^{-N s_4/{\hat s}} {\hat\sigma}(s_4)$ (1PI). Then the logarithms of $N$ in ${\hat \sigma}(N)$ exponentiate.

Consider the partonic process $f_1 +f_2 \rightarrow t+X$ where $X$ represents
the additional final-state particles apart from a produced top quark.
We proceed with the derivation of the resummed cross section by  
writing a factorized form for the moment-space 
parton-parton scattering cross section, infrared-regularized by $\epsilon$, 
$\sigma_{f_1 f_2 \rightarrow tX}(N, \epsilon)$, 
which factorizes as the hadronic cross section 
\beq
\sigma_{f_1 f_2 \rightarrow tX}(N, \epsilon)  
={\tilde \phi}_{f_1/f_1}(N,\mu_F,\epsilon)\; 
{\tilde \phi}_{f_2/f_2}(N,\mu_F,\epsilon) \;
{\hat \sigma}_{f_1 f_2 \rightarrow tX}(N,\mu_F,\mu_R) \, ,
\label{sig1}
\eeq
where the moments of $\phi$ are given by  
$\tilde{\phi}(N)=\int_0^1dx\; x^{N-1}\phi(x)$, and for 
simplicity we do not show the dependence of the cross section on 
kinematical variables.  
We factorize the initial-state collinear divergences 
into the parton distribution functions, $\phi$, 
which are expanded to the 
same order in $\alpha_s$ as the partonic cross section, 
and thus obtain the perturbative expansion for the
infrared-safe partonic short-distance function ${\hat \sigma}$.

The partonic function ${\hat \sigma}$ is still sensitive
to soft-gluon dynamics through its $N$ dependence.   
We then refactorize the moments of the cross section
as \cite{Kidonakis:1997gm}
\beqa
&&\sigma_{f_1 f_2\rightarrow tX}(N,\epsilon)
={\tilde\psi}_{f_1/f_1} \left(N,\mu_F,\epsilon \right) \;
{\tilde\psi}_{f_2/f_2} \left(N,\mu_F,\epsilon \right)  
\nonumber \\ && \hspace{-10mm} \times \; 
H_{IL}^{f_1 f_2\rightarrow tX} \left(\alpha_s(\mu_R)\right)\; 
{\tilde S}_{LI}^{f_1 f_2 \rightarrow tX} 
\left({M\over N \mu_F },\alpha_s(\mu_R) \right)\;
\prod_j  {\tilde J_j}\left (N,\mu_F,\epsilon \right)  
+{\cal O}(1/N) \, ,
\label{sig2}
\eeqa 
where $\psi$ are center-of-mass distributions that absorb 
the universal collinear singularities from the incoming partons, 
$H_{IL}$ is an $N$-independent function  
describing the hard-scattering, $S_{LI}$ is a soft gluon function
associated with non-collinear soft gluons, and $J$ are functions 
that absorb the collinear singularities from any massless partons  
in the final state. We note that the $J$ functions do not appear 
in the resummation for $t{\bar t}$ production but are needed in 
single top production via the $t$- and $s$-channels.

$H$ and $S$ are matrices in the space of the color structure of the 
hard scattering, with color indices $I$ and $L$.  
The hard-scattering function involves contributions from the 
amplitude of the process and its complex conjugate,
$H_{IL}=h_L^*\, h_I$.

Using Eqs. (\ref{sig1}) and (\ref{sig2}) we can write ${\hat{\sigma}}$
in terms of $H$, $S$, $J$, and the ratios $\psi/\phi$.
The constraint that the product of these functions must be
independent of the gauge and factorization scale results
in the exponentiation of logarithms of $N$ in the parton and soft functions 
\cite{Kidonakis:1997gm,Contopanagos:1996nh}. 

The soft matrix $S_{LI}$ depends on $N$ through the ratio $M/(N\mu_F)$,
and it requires renormalization as a composite operator.
However, in the product $H_{IL}\, S_{LI}$ the UV divergences of $S$ are
balanced by those of $H$.
$S_{LI}$ satisfies the
renormalization group equation~\cite{Kidonakis:1996aq,Kidonakis:1997gm}
\begin{equation}
\left(\mu {\partial \over \partial \mu}
+\beta(g_s){\partial \over \partial g_s}\right)\,S_{LI}
=-(\Gamma^\dagger_S)_{LB}S_{BI}-S_{LA}(\Gamma_S)_{AI}\, ,
\label{RGE}
\end{equation}
where $\beta(g_s)$ is the QCD beta function and $g_s^2=4\pi\alpha_s$.
$\Gamma_S$ is the soft anomalous dimension matrix, and it 
is calculated in the eikonal approximation 
by explicit renormalization of the soft function.
In a minimal subtraction renormalization scheme with
$\epsilon=4-n$, $\Gamma_S$ is given at one loop by
\begin{equation}
\Gamma_S^{(1)} (g_s)=-\frac{g_s}{2} \frac {\partial}{\partial g_s}
{\rm Res}_{\epsilon\rightarrow 0} Z_S (g_s, \epsilon) \, .
\label{GammaS1l}
\end{equation}
In processes with simple color structure $\Gamma_S$ is 
a $1\times 1$ matrix while in processes with complex 
color structure it is a non-trivial matrix in color exchange.
For quark-antiquark scattering into a top-antitop pair,  
$\Gamma_S$ is a $2\times 2$ matrix \cite{Kidonakis:1996aq,Kidonakis:1997gm};
for gluon-gluon fusion into a top-antitop pair it is a $3\times 3$ matrix 
\cite{Kidonakis:1997gm}.  

The exponentiation of logarithms of $N$ in $\psi/\phi$ and $J$ 
together with the solution of 
the renormalization group equation (\ref{RGE}), 
provide us with the complete expression for the resummed partonic cross 
section in moment space \cite{Kidonakis:1997gm}

\beqa
{\hat{\sigma}}^{res}(N) &=&   
\exp\left[ \sum_{i=1,2} E^{f_i}(N_i)\right] \; 
\exp\left[ \sum_j {E'}^{f_j}(N')\right] \;
\exp \left[\sum_{i=1,2} 2\int_{\mu_F}^{\sqrt{\hat s}} \frac{d\mu}{\mu}\;
\gamma_{i/i}\left({\tilde N}_i,\alpha_s(\mu)\right)\right] \;
\nonumber\\ && \hspace{-10mm} \times \,
{\rm Tr} \left \{H^{f_1 f_2 \rightarrow tX}\left(\alpha_s(\sqrt{\hat s})\right) \;
\exp \left[\int_{\sqrt{\hat s}}^{{\sqrt{\hat s}}/{\tilde N'}} 
\frac{d\mu}{\mu} \;
\Gamma_S^{\dagger \, f_1 f_2 \rightarrow tX}\left(\alpha_s(\mu)\right)\right] \;
{\tilde S^{f_1 f_2 \rightarrow tX}} \left(\alpha_s(\sqrt{\hat s}/{\tilde N'}) \right) \right. 
 \nonumber\\ && \quad \left.\times \,
\exp \left[\int_{\sqrt{\hat s}}^{{\sqrt{\hat s}}/{\tilde N'}} 
\frac{d\mu}{\mu}\; \Gamma_S^{f_1 f_2 \rightarrow tX}
\left(\alpha_s(\mu)\right)\right] \right\} \, .
\label{resHS}
\eeqa

In 1PI kinematics $N_i=N (-{\hat t}_i/M^2)$ for incoming partons $i$,
and  $N'=N ({\hat s}/M^2)$; here $M$ is any
chosen hard scale relevant to the process, such as the top quark mass.
In PIM kinematics $N_i=N'=N$. 
Also note that ${\tilde N}=N e^{\gamma_E}$, with $\gamma_E$ the Euler constant.

The first exponent in Eq. (\ref{resHS}) arises from the exponentiation
of logarithms of $N$ in the ratios $\psi/\phi$ (the sum $i=1,2$ is over 
incoming partons), 
and it is given in the $\overline{\rm MS}$ scheme by 
\beq
E^{f_i}(N_i)=
\int^1_0 dz \frac{z^{N_i-1}-1}{1-z}\;
\left \{\int_1^{(1-z)^2} \frac{d\lambda}{\lambda}
A_i\left(\alpha_s(\lambda \hat s)\right)
+D_i\left[\alpha_s((1-z)^2 \hat s)\right]\right\} \, ,
\label{Eexp}
\eeq
where $A$ \cite{Sterman:1986aj,Catani:1989ne} and $D$ 
\cite{Contopanagos:1996nh} have well-known perturbative expansions.

The second exponent in Eq. (\ref{resHS}) arises from the exponentiation of 
logarithms of $N$ in the functions $J_j$ for final-state particles 
(the sum $j$ is over outgoing quarks and gluons) 
\cite{Sterman:1986aj,Catani:1989ne}. It does not appear in 
$t{\bar t}$ production but is needed for single top quark production.

The third exponent in Eq. (\ref{resHS}) 
controls the factorization scale dependence of the cross 
section, and $\gamma_{i/i}$ is the moment-space 
anomalous dimension of the ${\overline {\rm MS}}$ density $\phi_{f_i/f_i}$.
We have
$\gamma_{i/i}=-A_i \ln {\tilde N_i} +\gamma_i$
with $\gamma_i$ the parton anomalous dimensions.
 
One can evaluate the anomalous dimensions and matching functions
appearing in the resummed cross section (\ref{resHS}) order-by-order in
perturbation theory.  The NNLL calculations mentioned at the beginning of the
section require the anomalous dimensions $A$ at three loops, all other
anomalous dimensions at two loops, and the hard and soft functions at NLO.
Obviously, every time a term is added to the anomalous dimensions a whole
tower of logarithms in the fixed-order expansion is resummed into the
exponent, which accounts for the nomenclature behind this re-organization of
the perturbative series.  A feature of the resummed cross section in Mellin
space is that it contains factors of $\alpha_s$ evaluated below the QCD scale
$\Lambda_{\rm QCD}$, and is thus subject to Landau-pole ambiguities related to
how to deal with this singularity. We will not discuss this issue in detail,
but one way of circumventing this problem is to instead construct fixed-order
expansions of (\ref{resHS}), a topic we return to in
Section~\ref{sec:NNLOapprox}.

\subsection{Resummation with SCET}
\label{sec:SCET}

We will explain the SCET approach to resummation using PIM kinematics
as an illustrative example \cite{Ahrens:2010zv}.  The same techniques
apply to 1PI kinematics after only small modifications
\cite{Ahrens:2011px}.  The starting point is the observation that
partonic cross sections near threshold factorize into a product of
a hard function, related to virtual corrections, and a soft function,
related to real emission in the soft limit \cite{Kidonakis:1997gm}.
We write this factorization, valid up to corrections of $\mathcal{O}(1-z)$,
as\footnote{The parametric
  scaling $M_{t\bar t}\sim m_t$ is assumed
  , which is valid as long as the top quarks are
not too highly boosted.}
\begin{align}
\label{eq:PIMfact}
  \frac{d\hat\sigma_{ij}(z,M_{t\bar t},m_t,\cos\theta,\mu_F)}{d M_{t\bar t} d\cos\theta} 
= {\rm Tr} \left[ \bm{H}_{ij}(M_{t\bar t},m_t,\cos\theta,\mu_F) \,
    \bm{S}_{ij}(\sqrt{\hat{s}}(1-z),m_t,\cos\theta,\mu_F) \right]  \, .
\end{align}  
The boldface indicates that the hard functions $\bm{H}_{ij}$ and soft
functions $\bm{S}_{ij}$ are matrices in color space. In SCET, the 
separation of the partonic cross sections into a product of 
hard and soft functions is achieved by a two-step matching 
procedure from QCD to the effective theory.
Hard fluctuations are integrated out in a  first step and the 
hard matching functions are identified as Wilson coefficients of
an operator built of soft and collinear fields. Soft fluctuations 
are integrated out in a second step and the soft matching functions
are the Wilson coefficient of a collinear operator, whose matrix 
element defines the PDFs.  Resummation of large logarithms 
is then performed by deriving and solving the RG equations for 
the two functions.  The main technical complication is that 
the RG equation for the soft function is non-local.
This is dealt with  using the Laplace-transform technique 
introduced in \cite{Becher:2006nr}, which observed that 
the  Laplace transformed functions 
\begin{align}
 \label{eq:stilde}
\tilde{\bm{s}}(L,M,m_t,\cos\theta,\mu) &= \frac{1}{\sqrt{\hat s}} \int_0^\infty d\omega
  \, \exp \left( -\frac{\omega}{e^{\gamma_E}\mu e^{L/2}} \right)
  \bm{S}(\omega,M,m_t,\cos\theta,\mu)\, , 
\end{align}
can be shown to obey local evolution equations whose solution can be
inverted back to momentum space analytically.

The final expression for the resummed partonic cross section in PIM 
kinematics is
\begin{align}
  \label{eq:MasterFormula}
  \frac{d\hat\sigma(z,M_{t\bar t},m_t,\cos\theta,\mu_F)}{d M_{t\bar t} d\cos\theta}  &= 
\exp \left [4a_{\gamma^{\phi}}(\mu_s,\mu_F) 
- 2a_{\Gamma}(\mu_s,\mu_F)\ln\frac{M_{t\bar t}^2}{\mu_s^2} \right]
  \nonumber
  \\
  &\hspace{-2em} \times {\rm Tr} \Bigg[ \bm{U}(M,m_t,\cos\theta,\mu_h,\mu_s) \,
  \bm{H}(M,m_t,\cos\theta,\mu_h) \, \bm{U}^\dagger(M,m_t,\cos \theta,\mu_h,\mu_s)
  \nonumber
  \\
  &\hspace{-2em} \times \tilde{\bm{s}}
  \left(\partial_\eta,M,m_t,\cos\theta,\mu_s\right) \Bigg]
  \frac{e^{-2\gamma_E \eta}}{\Gamma(2\eta)}
  \frac{1}{(1-z)}\left(\frac{2E^{\rm PIM}_s(z)}{\mu_s}\right)^{2\eta}\bigg|_{\eta =2a_{\Gamma}(\mu_s,\mu_F)  } \, .
\end{align}
In addition to the hard and soft matching functions, the formula
contains factors related to the RG evolution from the matching scales
$\mu_h$ and $\mu_s$ to the factorization scale $\mu_F$. These RG
factors are given in terms of integrals over anomalous dimensions;
their exact definitions can be found in \cite{Ahrens:2010zv}.  For
values $\mu_s<\mu_F$ the parameter $\eta<0$, and one must use a
subtraction at $z=1$ and analytic continuation to express integrals
over $z$ in terms of plus distributions.
Formula~(\ref{eq:MasterFormula}) can be evaluated order-by-order in
RG-improved perturbation theory, using the standard counting $\ln
\mu_h/\mu_s\sim \ln (1-z)\sim 1/\alpha_s$.  The current state of the
art is NNLL \cite{Ahrens:2010zv}, which roughly speaking requires the
soft anomalous dimension at two loops (as obtained in
\cite{Ferroglia:2009ep, Ferroglia:2009ii}) and the hard and soft
matching functions to NLO.  

Note that the logarithmic corrections at the soft scale are generated
by derivatives with respect to $\eta$ acting on the factors of
$(2E^{\rm PIM}_s(z)/\mu_s)^{2\eta}$.  In the SCET calculation, it is
natural to use $2E^{\rm PIM}_s(z)=M_{t\bar t}(1-z)/\sqrt{z}$, which is
the energy of the soft radiation in the partonic center-of-mass frame.
We will refer to such a choice as the $\scetpim$ scheme.  An equally
appropriate method, taken in all earlier literature on soft-gluon
resummation, is to use the expansion of the energy in the $z\to 1$
limit, and we will refer to such a choice as the PIM scheme--the
situation is summarized in Table~\ref{tab:scetschemes}.  The
difference between the two schemes involves power-suppressed terms of
the form $\ln^nz/(1-z)$ when the formula is re-expanded in
fixed order. Such terms appear indeed naturally in the fixed-order
calculations, as observed in the case of SCET applications to
Drell-Yan \cite{Becher:2007ty} and Higgs production
\cite{Ahrens:2008qu,Ahrens:2008nc}.  At NLO it can be shown
numerically that keeping such terms in the factorized cross section
(\ref{eq:PIMfact}) reduces the size of the remaining power-suppressed
corrections which vanish in the $z\to 1$ limit.  We will come back to
this point in the discussion of approximate NNLO formulas in 
Section~\ref{sec:NNLOapprox}.

The methods of factorization and resummation described above for the
case of PIM kinematics generalizes to that of 1PI kinematics in a
straightforward way.  In fact, the structure of the resummed formula
is exactly as in (\ref{eq:MasterFormula}).  The main difference is
that the expression for the soft function changes, since it is derived
by integrating over a different region of the phase space.  The hard
function, on the other hand, is the same, as is the form of the
evolution equations. An analysis to NNLL, which required the
calculation of the NLO corrections to the soft function, was performed
in \cite{Ahrens:2011mw}. Similarly to the case of PIM kinematics,
logarithmic corrections in 1PI kinematics are generated by derivatives
acting on a factor of $(2E^{\rm 1PI}_s(s_4)/\mu_s)^{2\eta}$
and one has the choice between identifying this
factor with the exact energy of soft-gluon radiation in the ${\bar t}$ rest
frame or  its expansion in the $s_4\to 0$ limit. The two schemes
defined in this way are summarized in Table~\ref{tab:scetschemes}.
The difference between the $\scetopi$ and 1PI schemes
involves terms of the form $\ln^n(1+s_4/m_t^2)/s_4$ when expanded in
fixed order.  Even more so than for the case of PIM kinematics, these
subleading terms make an important numerical difference.

The SCET approach to soft-gluon resummation can also be applied to the
total cross section in the $\beta \to 0$ limit.  In this limit one can perform
soft gluon resummation through the same techniques as in PIM and 1PI
kinematics.  However, such an approach is insufficient because one must also
take into account Coulomb singularities of the form $\ln^m\beta/\beta^n$,
and a joint resummation of the two effects is more complicated than soft gluon
resummation alone. This issue was solved in \cite{Beneke:2009rj,
Beneke:2010da}, which used techniques in SCET and non-relativistic QCD to
perform a combined resummation of Coulomb and soft-gluon effects to NNLL
order.

From a technical standpoint, resummation within the framework of SCET
is very similar to the Mellin-space resummation described in the
previous section.  In fact, the resummed formulas are completely
equivalent if consistently re-expanded to any given order in
$\alpha_s$.\footnote{This was shown in detail in
  \cite{Becher:2006nr,Becher:2006mr}, and later in many other cases,
  including heavy-quark production in the production threshold limit
  \cite{Beneke:2009rj}.  The same techniques apply to PIM and 1PI
  kinematics but an analysis has not yet appeared in the literature.}
Other than the fact that the SCET results apply directly in momentum
space, the main difference between the two approaches is the way in
which the soft matching scale $\mu_s$ is chosen, and the reasoning
underlying this choice. In the SCET approach, one argues that the
appearance of a well-separated soft scale at the level of hadronic
cross sections is a dynamical effect due to the sharp fall-off of the
PDFs away from the partonic threshold region \cite{Becher:2007ty}.  To
determine its numerical value, one studies the corrections arising
from the soft function to the differential hadronic cross section as a
function of $\mu_s$ in fixed order, and finds the value at which these
corrections are minimized.  This numerical value is a function of the
kinematic variables observed in the differential cross section (for
instance in PIM kinematics of the invariant mass $M_{t\bar t}$), and
is interpreted at the scale at which the soft function is free of
large logarithmic corrections.  Such a choice eliminates the
Landau-pole ambiguity, but also implies that the SCET resummation
exponentiates a subset of the higher-order plus distributions along
with logarithms of the numerical ratio $\mu_s/\mu_F$; a more complete
discussion can be found in the Appendix of \cite{Ahrens:2011mw}.

%%%%%%%%%%%%%%%%%%%%%5
\begin{table}
\begin{center}
\begin{tabular}{|l|l||l|l|}
\hline
 $\scetpim$ & $2E^{\rm PIM}_s(z) = M_{t\bar t} (1-z)/\sqrt{z}$ &
$\scetopi$ & $2E^{\rm 1PI}_s(s_4) =  s_4/\sqrt{m_t^2+s_4}$ \\
PIM& $2E^{\rm PIM}_s(z) =  M_{t\bar t}(1-z)$ &
1PI& $2E^{\rm 1PI}_s(s_4)=  s_4/m_t$ 
\\
\hline
\end{tabular}
\end{center}
\caption{\label{tab:scetschemes} The values of the 
parameter $E_s$ which define different calculational schemes used
in this review.}
\end{table}
%%%%%%%%%%%%%%%%%%%%%%%%%%%%%%%%%%%

\subsection{Approximate NNLO formulas}
\label{sec:NNLOapprox}
In the previous two sections we focused on all-orders resummation formulas.
The basic idea of such formulas is to use properties of real radiation in the
soft limit to calculate an infinite set of logarithmic corrections to the
partonic cross section in terms of a smaller group of objects such as
anomalous dimensions and matching functions.  An alternative is to use the
same formalism to determine only the logarithmic corrections to a certain
accuracy in the fixed-order expansion.  This is a useful approximation if it
can be argued that such logarithmic terms capture the dominant corrections at
a given order, and also that the higher-order logarithmic corrections are not
so large as to spoil the convergence of the fixed-order expansion. A benefit 
of using such an expansion is that the Landau-pole singularities in the
Mellin-space formalism are absent, and there is no need to introduce a 
dynamically generated numerical soft scale in the SCET approach. In this
section we explain the structure of  approximate NNLO formulas derived
from NNLL resummation within the different types of soft limits.  We point out
that such formulas are subject to a number of ambiguities in the treatment of
subleading terms in the soft limit, and discuss how these are dealt with in
the literature.

We begin with a discussion of 1PI and PIM kinematics.  The approximate 
NNLO formulas in these cases are applicable to the differential 
distributions shown in Table~\ref{tab:softlimits}, but to compare 
with the production threshold limit we will work at the level 
of the total partonic cross section.  In PIM kinematics, one can 
write the general form of the NNLO corrections to the partonic 
scaling functions (\ref{eq:fexp}) as 
(suppressing the labels for the $q\bar q$ and $gg$ channels)
\begin{align}
\label{eq:PIMapprox}
f^{(2)}(\beta,\mu)=\int  d\cos\theta dz\left[\sum\limits_{n=0}^3 
D^{\rm PIM}_n\left[ \frac{\ln^n(1-z)}{1-z} \right]_+ + C^{\rm PIM} \delta(1-z) 
+R^{\rm PIM}(z) \right] \, ,
\end{align}
while in 1PI kinematics, one can write
\begin{align}
\label{eq:1PIapprox}
f^{(2)}(\beta,\mu)& = \int d\hat{t}_1 ds_4 \left[\sum\limits_{n=0}^3 
D_n^{\rm 1PI} \left[ \frac{\ln^n(s_4/m_t^2)}{s_4} \right]_+ +
 C^{\rm 1PI} \delta(s_4) 
+R^{\rm 1PI}(s_4) \right] \, .
\end{align}
The $D_i$, $C$, and $R$ coefficients are functions of the variables
$\hat{s},\hat{t}_1,\hat{u}_1,m_t,\mu$, so one evaluates the formulas
above after an appropriate change of variables.  The definition is
such that the $R$ coefficients are regular in the limit $z\to 1$ or
$s_4 \to 0$.  Soft gluon resummation at NNLL accuracy determines all
of the $D_i$ coefficients in the limit $z\to 1$ or $s_4\to 0$, and
also the $\mu$-dependent piece of the delta-function coefficient $C$
\cite{Kidonakis:2010dk, Ahrens:2011mw, Ahrens:2009uz}.  To determine
the $\mu$-independent piece of the $\delta$-function coefficient would
require the calculation of the soft and hard matching functions at
NNLO order.  The regular coefficients $R$ are not determined by the
soft gluon resummation in the form discussed here and can only be
obtained through a complete NNLO calculation.

An alternative to the 1PI and PIM results is to approximate the NNLO
corrections through the singular terms in the production threshold
limit $\beta \to 0$.  The complete answer was first obtained in
\cite{Beneke:2009ye}.  To illustrate its structure, we decompose the
NNLO correction to the scaling functions (\ref{eq:fexp}) as
\begin{align}
  \label{eq:f2exp}
  f_{ij}^{(2)} &= 
    f_{ij}^{(2,0)} + f_{ij}^{(2,1)} \ln\left(\frac{\mu_F^2}{m_t^2}\right) + f_{ij}^{(2,2)}
    \ln^2\left(\frac{\mu_F^2}{m_t^2}\right)  .
\end{align}
Up to pieces which are regular in the $\beta \to 0$ limit, the 
scale-independent corrections read  \cite{Beneke:2009ye} 
\begin{align}
  \label{eq:betaexpans}
  f_{q\bar q}^{(2,0)} &= \frac{1}{(16\pi^2)^2} \frac{\pi\beta}{9} 
\Big[\frac{3.60774}{\beta^2}
+\frac{1}{\beta}\Big(-140.368\lb2+32.106\lb1+3.95105\Big) \nonumber \\ &
+910.222\lb4-1315.53\lb3+592.292\lb2+528.557\lb1 \Big] + \ldots , \nonumber
  \\
  f_{gg}^{(2,0)} &= \frac{1}{(16\pi^2)^2} \frac{7\pi\beta}{192} 
\Big[ \frac{68.5471}{\beta^2}
+\frac{1}{\beta}\Big(496.3\lb2+321.137\lb1-8.62261\Big)\nonumber \\ &
+4608\lb4-1894.91\lb3-912.349\lb2+2456.74\lb1 \Big] + \ldots .
\end{align}
Earlier results, which give very similar numerical results for the
total cross section although the analytic structure is incorrect, were
obtained in \cite{Moch:2008qy}.  In that paper the coefficients
multiplying the $\mu$-dependent logarithms in (\ref{eq:f2exp}), exact
in $\beta$, were also presented.

In Section~\ref{sec:totalcs} we will compare results for the total top-pair
production cross section based on three approximate NNLO calculations
available in the literature: the 1PI threshold as implemented in
\cite{Kidonakis:2010dk}, the 1PI and PIM thresholds as implemented in the
computer program from \cite{Ahrens:2011mw}, and the production threshold as
implemented in the HATHOR program \cite{Aliev:2010zk}.  Since
\cite{Kidonakis:2010dk} and \cite{Ahrens:2011mw} both use 1PI kinematics and
therefore differ only through implementations, it is important to clarify the
sources of difference between the 1PI results from \cite{Kidonakis:2010dk} and
those from the SCET-based approach in \cite{Ahrens:2011mw}.  Here there are
several issues to consider.  First, the numerical results depend on how the
relation $\hat{s}+\hat{u}_1+\hat{t_1}=0$ is used in the plus-distribution
coefficients (\ref{eq:1PIapprox}) before numerical integration, and the two
papers in general use different re-writings which are equivalent in the limit
$s_4\to 0$.  Second, even if this relation were used in the same way, there
would still be differences in the analytic structure of the approximate NNLO
formulas.  These definitely affect the $\mu$-dependent part of the
$\delta$-function terms, but possibly also the $D_0$ and $\mu$-independent
part of the $\delta$-function terms.  The expressions for these coefficients
used in \cite{Kidonakis:2010dk} can be obtained from
\cite{Kidonakis:2003qe,Kidonakis:2005kz,Kidonakis:2010dk}, while those from
\cite{Ahrens:2011mw} were specified in that paper and given in the form of a
computer program.  Finally, the authors of \cite{Ahrens:2011mw} prefer the
$\scetopi$ (and $\scetpim$) implementation of the approximate NNLO formulas
(see Table~\ref{tab:scetschemes} and the discussion in
Section~\ref{sec:SCET}), where a certain set of subleading terms appearing
naturally within the SCET formalism is kept in the regular coefficients $R$.
On the other hand, \cite{Kidonakis:2010dk} uses damping factors 
$2m_t/\sqrt{\hat s}$ in the
NNLO soft-gluon corrections to the total cross section (for the $p_T$
distributions a damping factor $2m_{\perp}/\sqrt{\hat s}$ is used) in order to
reduce the contribution from kinematical regions further away from threshold, 
and thus improve the 1PI approximation.  For the total cross section, the
damping factors lead to differences between \cite{Kidonakis:2010dk} and
\cite{Ahrens:2011mw} that are actually not subleading in the $s_4\to 0$ limit
of the NNLL calculation.  At NLO, the agreement between the exact
result for the total cross section and its threshold approximation is
improved in both the $\scetopi$ scheme and the 1PI scheme with damping factors
compared to the pure 1PI expansion. However, at NNLO these different schemes 
produce rather different results, and the numerical effect of 
using damping factors versus including subleading SCET terms is greater than 
that from the other differences in implementation mentioned earlier in this 
paragraph. More details can be found in the numerical
studies in \cite{Kidonakis:2010dk} and \cite{Ahrens:2011mw}, and also
in the comparison in the next section.

\section{Total top-pair production cross section}
\label{sec:totalcs}
The most basic quantity in top-quark pair production is the total
inclusive cross section.  In this section we collect numerical results
for this quantity based on different levels of perturbative accuracy
and discuss the associated uncertainties.  We focus on a comparison of
NLO and approximate NNLO predictions derived from NNLL resummation in
three types of soft limits: production threshold, PIM, and 1PI.

To provide a QCD prediction for the total cross section at a
particular collider, one must decide on two things.  First, the value
of input parameters such as the top-quark mass and a PDF set, and
second, a procedure for estimating the theoretical uncertainties
associated with the perturbative corrections beyond the accuracy of
the calculation.  For the NLO calculation, one typically uses scale
variations to estimate the uncertainties related to uncalculated
corrections at NNLO and beyond.  For the approximate NNLO calculations
the uncalculated terms include the pieces of the NNLO correction which
are subleading in the limit in which the formula is derived.  The
situation is thus slightly different from a full fixed-order
calculation and there is no conventional way for estimating
theoretical uncertainties: some authors use the standard method of
scale variations, some use a more complicated procedure. We describe
the different approaches in conjunction with the results below.

The results for the cross section in the pole scheme are summarized
in Table~\ref{tab:cspole}. In addition to the NLO results, we use
the following implementation of approximate NNLO predictions:
\begin{itemize}
\item Production threshold results as obtained by the HATHOR program
  \cite{Aliev:2010zk} with the default settings. 
\item 1PI results as obtained in \cite{Kidonakis:2010dk}.
\item $\scetopi$ and $\scetpim$ results as obtained in
  \cite{Ahrens:2011mw}. These are combined into a final result for the
  cross section using the procedure and computer program presented in
  \cite{Ahrens:2011px}.
\end{itemize}
By default, we have set $\mu=\mu_R=\mu_F
= m_t$, with $m_t=173$~GeV.  We display NLO results using 
MSTW2008 NLO PDFs, while for approximate NNLO results we use
MSTW2008 NNLO PDFs \cite{Martin:2009iq}.  Uncertainties in the HATHOR and 1PI results from
\cite{Kidonakis:2010dk} are estimated by varying $\mu$ up and down by
a factor of two{\footnote{Note that independent variations of $\mu_F$ and 
$\mu_R$ in \cite{Kidonakis:2010dk} do not increase the uncertainty for LHC 
energies.}}, while uncertainties in \cite{Ahrens:2011px} are
estimated by independent variations of $\mu_R$ and $\mu_F$ by factors
of two, along with a scan over the values of the cross section in PIM
and 1PI kinematics.

\begin{table}[h]
\begin{center}
  \begin{tabular}{c|c|c}
 & Tevatron &  LHC (7\,TeV) \\ \hline
 NLO  & 
  6.74{\footnotesize $^{+0.36}_{-0.76}$}{\footnotesize
    $^{+0.37}_{-0.24}$}  &
  160{\footnotesize $^{+20}_{-21}$}{\footnotesize $^{+8}_{-9}$}\\
Aliev et. al.  \cite{Aliev:2010zk} & 
  7.13{\footnotesize $^{+0.31}_{-0.39}$}{\footnotesize $^{+0.36}_{-0.26}$} &
  164{\footnotesize $^{+3}_{-9}$}{\footnotesize $^{+9}_{-9}$} 
\\
Kidonakis \cite{Kidonakis:2010dk} & 
  7.08{\footnotesize $^{+0.00}_{-0.24}$}{\footnotesize $^{+0.36}_{-0.24}$} &
  163{\footnotesize $^{+7}_{-5}$}{\footnotesize $^{+9}_{-9}$} 
\\ 
Ahrens et. al. \cite{Ahrens:2011px} &
 $6.65${\footnotesize$ ^{+0.08+0.33}_{-0.41-0.24}$} &
 $156${\footnotesize$ ^{+8+8}_{-9-9}$} 
\end{tabular}
\end{center}
\vspace{-2mm}
\caption{\label{tab:cspole} Results for the total cross section in pb
  at NLO and within the various NNLO approximations. The first uncertainty 
  is related to perturbative uncertainties, and the second is the PDF error 
  using the MSTW2008 PDF sets \cite{Martin:2009iq} at 90\% CL.  }
\end{table}

An examination of the numbers in the table reveals the following
features.  First, the perturbative uncertainties in the NLO result are
on the order of 20\% at both the Tevatron and the LHC.  This is a bit
larger than the PDF uncertainty in both cases, although especially at
the LHC one may obtain rather different results with other PDF sets,
we refer the reader to \cite{Watt:2011kp} for a recent discussion of
this issue.  Second, the perturbative uncertainties in the approximate
NNLO results as obtained through the individual calculations are
invariably smaller than at NLO--depending on the implementation, the
uncertainties are reduced by a factor of roughly two to three, and are
thus under the PDF uncertainties.  At the LHC, the different NNLO
approximations are in relatively good agreement, though the cross
section of \cite{Ahrens:2011px} is somewhat smaller than in
\cite{Kidonakis:2010dk, Aliev:2010zk}. At the Tevatron, on the other
hand, the results from \cite{Kidonakis:2010dk, Aliev:2010zk} are
significantly larger than those from \cite{Ahrens:2011px}.  In fact,
the range of values spanned by the three different approximate NNLO
results at the Tevatron is about as large as that spanned by the NLO
calculation.  Given the discrepancy, one is faced with the choice of
estimating the theory uncertainties through the NLO calculation, with
the spread of approximate NNLO values from the three different
calculations, or by a particular NNLO approximation alone.  The
authors of \cite{Ahrens:2011px, Kidonakis:2010dk, Aliev:2010zk} all
give arguments in favor of their particular implementation of
soft-gluon resummation, but it is beyond the scope of the review to
properly summarize them.  Instead, we describe very briefly the 
reasons behind the differences and explain the points which must 
be addressed to argue for one implementation over the others.

The most important discriminator between the approaches is of course
the soft limit in which the resummation is performed.  As mentioned
before, the production-threshold limit $\beta \to 0$ is actually a
special case of the 1PI and PIM thresholds.  At the level of the
approximate NNLO formulas, the production threshold results can be
obtained from the 1PI and PIM results by re-expanding them in that
limit, up to differences related to Coulomb terms (of the
form $\ln^m\beta/\beta^n$, see \cite{Ahrens:2010zv} for explicit
results) which turn out to be very small numerically
\cite{Kidonakis:2008mu, Ahrens:2010zv}, particularly at the Tevatron.
Therefore, in cases where the production threshold limit differs
substantially from the 1PI and PIM cases, it can be concluded that
subleading terms in $\beta$ are not generically small, so arguing for
the HATHOR results requires explaining why power corrections to the
partonic cross sections do not inherit this generic property and
furthermore why the subleading contributions included in the 1PI and
PIM results are non-physical.  Several differences between the 1PI
results of \cite{Kidonakis:2010dk} and the SCET-based results of
\cite{Ahrens:2011px} were described in Section~\ref{sec:NNLOapprox}.
The numerical importance of various differences is not obvious from
the table, since \cite{Ahrens:2011px} uses a scan over 1PI and PIM
kinematics as an estimate of power corrections, but a more detailed
study reveals that the dominant effects are the damping factors in
\cite{Kidonakis:2010dk} or the subleading terms in the $\scetopi$ scheme
in \cite{Ahrens:2011px}, and less so the differences in the
analytic structure described in Section~\ref{sec:NNLOapprox}.
Favoring the results of either \cite{Kidonakis:2010dk} or
\cite{Ahrens:2011px} thus requires arguments as to why the 1PI results
with damping factors or the combination of the SCET implementations
of the 1PI and PIM results gives a more reliable estimate of the cross
section and its uncertainties.

\begin{figure}[t]
\begin{center}
\psfrag{f}[1]{{\scriptsize CDF  \cite{CDFtt}  }}
\psfrag{e}[1]{{\scriptsize D0  \cite{Abazov:2011cq}  }}
\psfrag{d}[1]{{\scriptsize NLO}}
\psfrag{c}[1]{{\scriptsize Aliev et. al. \cite{Aliev:2010zk}}}
\psfrag{b}[1]{{\scriptsize Kidonakis \cite{Kidonakis:2010dk}}}
\psfrag{a}[1]{{\scriptsize Ahrens et. al. \cite{Ahrens:2011px}}}
\includegraphics[width=0.43\textwidth]{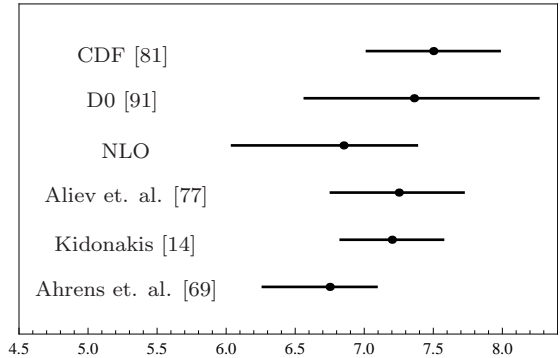} 
\caption{Experimental and theoretical values of the 
$t{\bar t}$ production cross section in pb at the Tevatron,
assuming a top-quark mass of 172.5~GeV.
The central values are indicated by the dots, the horizontal lines in the
experimental results represent uncertainties from a combination of 
statistical and systematic errors, and the horizontal lines in the 
theory results are the perturbative and PDF uncertainties added in quadrature.}
\label{fig:TEVCS}
\end{center}
\end{figure}

\begin{figure}[t]
\begin{center}
\psfrag{f}[1]{{\scriptsize ATLAS \cite{ATLAStt-2}  }}
\psfrag{e}[1]{{\scriptsize CMS \cite{Chatrchyan:2011nb} }}
\psfrag{d}[1]{{\scriptsize NLO}}
\psfrag{c}[1]{{\scriptsize Aliev et. al. \cite{Aliev:2010zk}}}
\psfrag{b}[1]{{\scriptsize Kidonakis \cite{Kidonakis:2010dk}}}
\psfrag{a}[1]{{\scriptsize Ahrens et. al. \cite{Ahrens:2011px}}}
\includegraphics[width=0.43\textwidth]{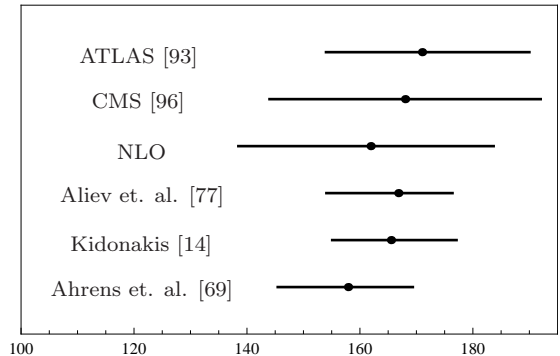} 
\caption{Experimental and theoretical values of the 
$t{\bar t}$ production cross section in pb at the LHC with $\sqrt{s}=7$~TeV,
assuming a top-quark mass of 172.5~GeV.
The central values are indicated by the dots, the horizontal lines in the
experimental results represent uncertainties from a combination of 
statistical and systematic errors, and the horizontal lines in the 
theory results are the perturbative and PDF uncertainties added in quadrature.}
\label{fig:LHCCS}
\end{center}
\end{figure}

Recent measurements of the production cross section have been made at
the Tevatron CDF
\cite{CDFtt,Aaltonen:2010pe,Aaltonen:2010bs,Aaltonen:2010se,Aaltonen:2010hza}
and D0
\cite{Abazov:2008gc,Abazov:2009ae,Abazov:2009ss,Abazov:2010pa,Abazov:2011mi,Abazov:2011cq}
experiments, and the LHC ATLAS \cite{ATLAStt-1,ATLAStt-2,ATLAStt-3}
and CMS \cite{Khachatryan:2010ez,Chatrchyan:2011nb,Chatrchyan:2011ew}
experiments.  An especially useful method for comparing with the
theory predictions is to show both as a function of the top-quark
mass.  Then the regions of overlap give a value of the top-quark mass
as determined from the production cross section.  While the errors in
the top-quark mass extracted in this way are larger than those from
kinematic distributions in top-quark decay, an advantage is that the
theory calculations are carried out in a well-defined
renormalization-scheme for the top-quark mass.  The results of such 
analyses carried out through measurements of the lepton + jets channel
by the D0 collaboration at the Tevatron appear in
\cite{Abazov:2008gc,Abazov:2011mi,Abazov:2011pt}.  
An analogous study has been performed by the ATLAS
collaboration at the LHC in \cite{ATLAS-mt}. 
It is also interesting to perform the analysis
using short-distance masses such as the $\msbar$
\cite{Langenfeld:2009wd,Ahrens:2011px} 
or threshold masses \cite{Ahrens:2011px};
results in the $\msbar$ scheme extracted from Tevatron data can be
found in \cite{Abazov:2011pt}. In Figures~\ref{fig:TEVCS}
and~\ref{fig:LHCCS} we compare recent experimental measurements
assuming a pole mass of $m_t=172.5$~GeV with the NLO and approximate
NNLO predictions, using the same method of estimating perturbative
uncertainties as described above and used in Table~\ref{tab:cspole}.
Although at the Tevatron the central values for an assumed top-quark
mass of $172.5$~GeV favor the higher values of the cross section
predicted by \cite{Kidonakis:2010dk, Aliev:2010zk}, the lower values
of $m_t$ extracted through the  NLO and approximate NNLO calculations from
\cite{Ahrens:2011px} are still within the world average \cite{Abazov:2011mi}.

\section{Top quark differential distributions}
\label{sec:diffcs}
We now turn to calculations of differential cross sections.  To keep
in line with the spirit of this review we cover only a few cases: the
$t\bar t$ invariant mass distribution and the top-quark $p_T$ and
rapidity distributions. Results for the
pair invariant mass distribution using soft gluon resummation at
NLO+NNLL or approximate NNLO exist from the calculations of
\cite{Ahrens:2010zv} within PIM kinematics, while those for the $p_T$
and rapidity distributions from the calculations of
\cite{Kidonakis:2010dk, Ahrens:2011mw, Kidonakis:2011zn} within 1PI
kinematics.  

\begin{figure}
\begin{center}
\begin{tabular}{lr}
\psfrag{x}[][][1]{$M_{t\bar t}$ [GeV]}
\psfrag{y}[][][1][90]{$d\sigma/dM_{t\bar t}$ [fb/GeV]}
\psfrag{z}[][][0.85]{$\sqrt{s}=1.96$\,TeV}
\includegraphics[width=0.44\textwidth]{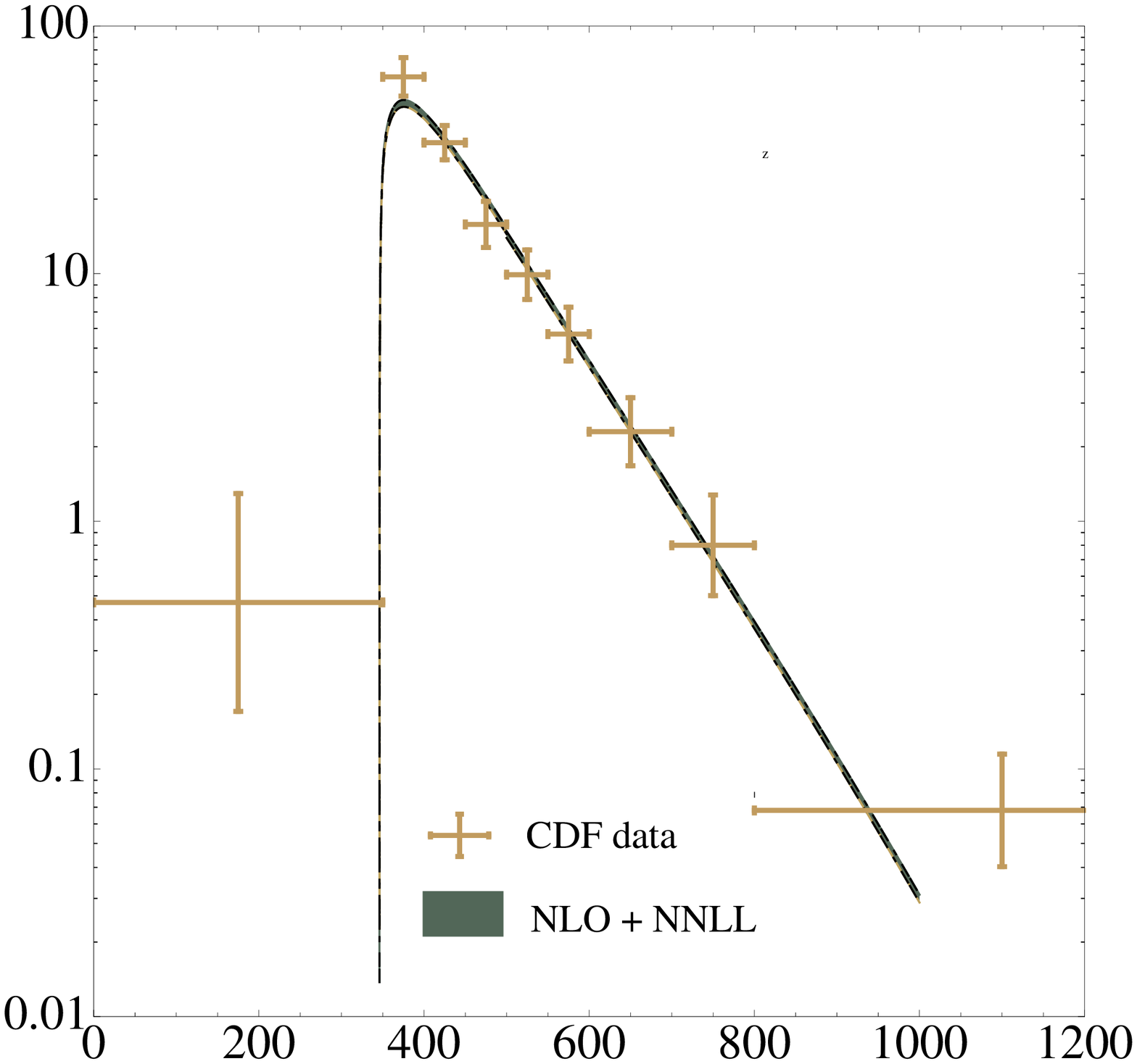}
\end{tabular}
\end{center}
\vspace{-2mm}
\caption{\label{fig:cdf-compare} Comparison of 
NLO+NNLL results \cite{Ahrens:2010zv}
for the invariant mass spectrum with CDF measurements 
\cite{Aaltonen:2009iz}.}
\end{figure}

We first discuss the pair invariant-mass distribution.  This is an 
important quantity for phenomenology.  It is sensitive to
searches for new physics through narrow $s$-channel resonances, which
would show up as bumps in the distribution, while within the SM itself
moments of the distribution give additional information about the
top-quark mass \cite{Frederix:2007gi}.  Moreover, recent measurements
of the forward-backward asymmetry by the CDF collaboration at high
values of the pair-invariant mass are much larger than SM predictions,
and an important constraint on new physics models which could produce
such an excess is the level of agreement of the invariant-mass
distribution with the SM calculations.  At the Tevatron, this level of
agreement is quite good, which we illustrate through the comparison of
the CDF result with the NLO+NNLL calculation \cite{Ahrens:2010zv} in
Figure~\ref{fig:cdf-compare}.  Obviously, there are
no narrow resonances in the experimental distribution at the Tevatron,
and measurements at the LHC will extend this search to higher values
of the invariant mass. It is worth emphasizing that at very large
values of the invariant mass the theory calculation is more
complicated.  This is mainly because the top quarks are highly boosted
at very high invariant mass, and at some point one must apply an
effective theory appropriate for the parametric scaling $m_t\ll M_{t\bar
  t}$.  Also, for very precise predictions, one must account for
electroweak corrections, since these contain Sudakov logarithms that
grow with the invariant mass \cite{Bernreuther:2006vg, Kuhn:2006vh}.

\begin{figure}
\begin{center}
\includegraphics[width=10cm]{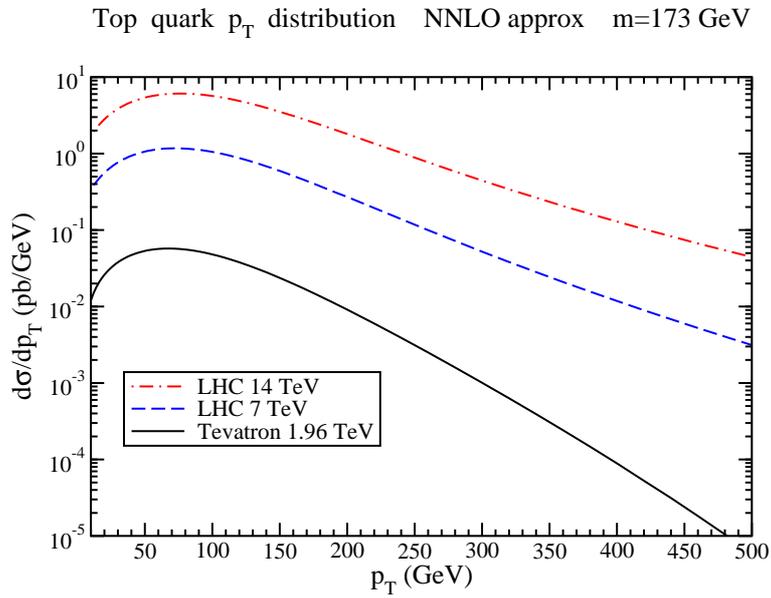}
\caption{The top quark transverse momentum distribution at the LHC and the 
Tevatron from \cite{Kidonakis:2010dk}.}
\label{ptlogplot}
\end{center}
\end{figure}

\begin{figure}
\begin{center}
\begin{tabular}{lr}
\psfrag{y}[][][1][90]{$d\sigma/dp_T$ [fb/GeV]}
\psfrag{x}[]{$p_T\,\rm{[GeV]}$}
\psfrag{z}[][][0.85]{$\sqrt{s}=1.96$\,TeV}
\includegraphics[width=0.50\textwidth]{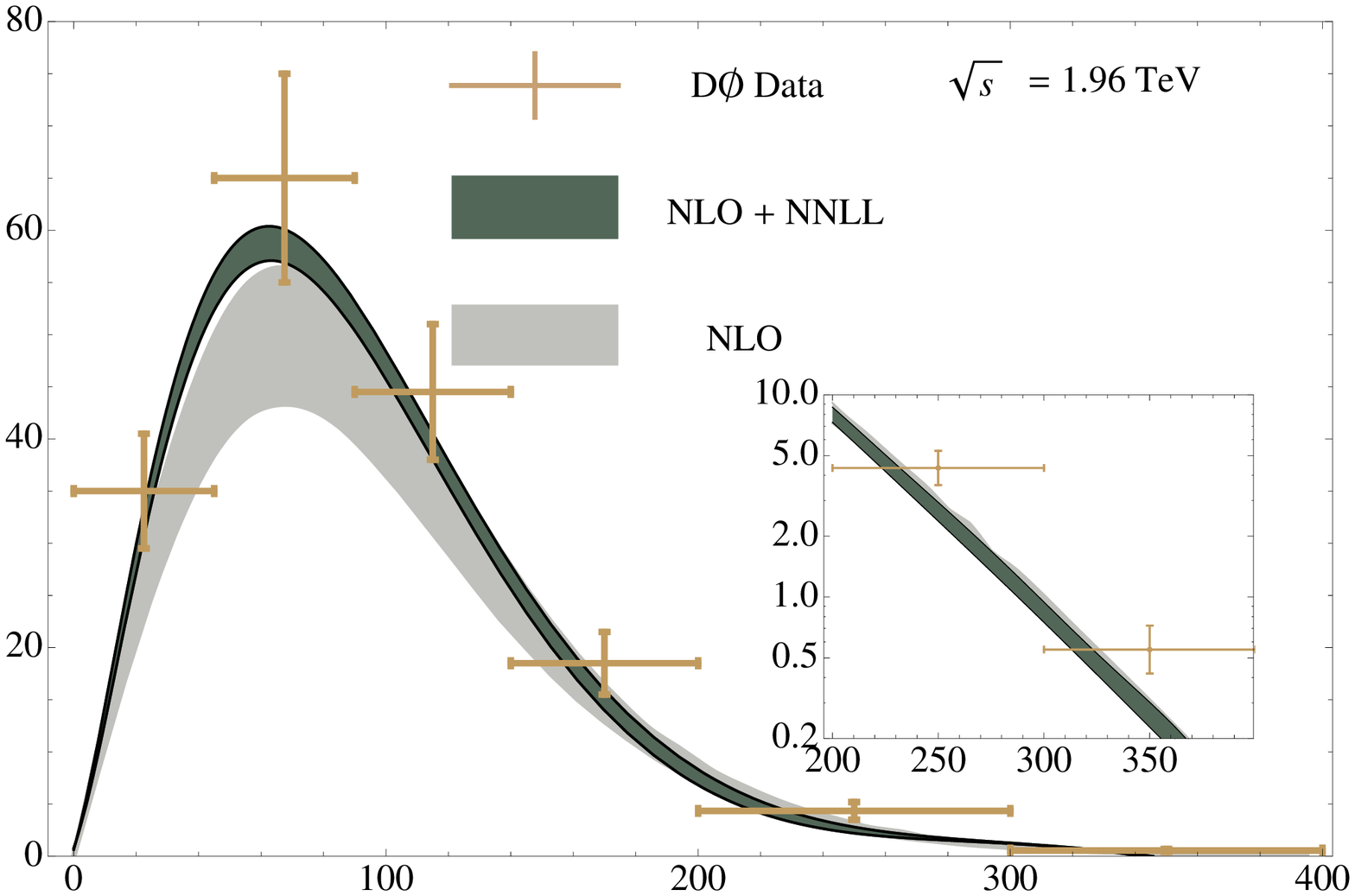}
\end{tabular}
\end{center}
\vspace{-2mm}
\caption{\label{fig:p_t-data} Comparison of NLO+NNLL
results \cite{Ahrens:2011mw} for the transverse-momentum
distributions with D0 measurements \cite{Abazov:2010js}. 
}
\end{figure}

Next, we discuss the top-quark transverse-momentum 
distribution.  Results for the $p_T$
distributions at approximate NNLO at the Tevatron and the LHC
with two collider energies are shown in Figure~\ref{ptlogplot}, based
on the calculations of \cite{Kidonakis:2010dk}. 
These results use MSTW2008 NNLO PDFs with $\mu_F=\mu_R=m_t=173$ GeV.  
\begin{figure}
\begin{center}
\includegraphics[width=10cm]{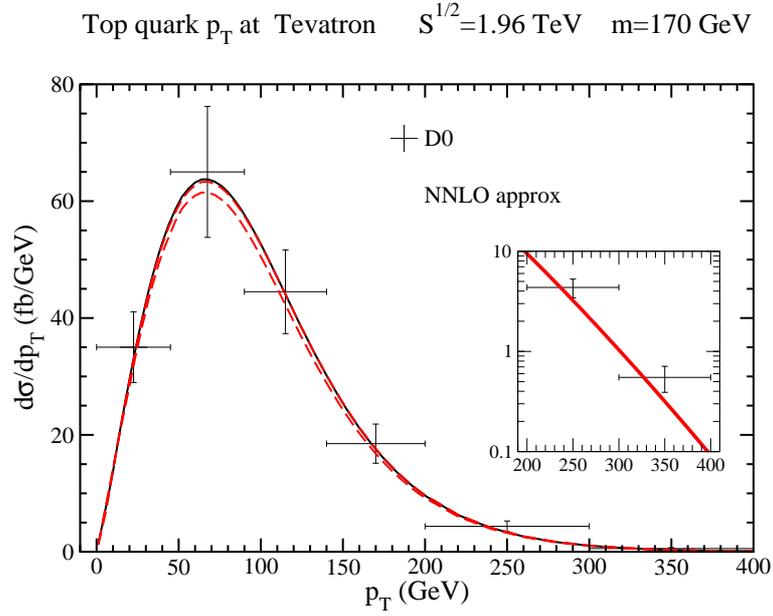}
\caption{The top quark transverse momentum distribution at the Tevatron 
from \cite{Kidonakis:2010dk} with $m_t=170$ GeV and $\mu=m_t/2,m_t,2m_t$ compared with 
D0 measurements \cite{Abazov:2010js}.}
\label{ptD0tevplot}
\end{center}
\end{figure}
The D0 collaboration in Ref. \cite{Abazov:2010js} has published measurements
of the top quark $p_T$ distribution and compared with various theoretical
results assuming $m_t=170$ GeV.  We show a comparison of the D0 
measurements with
the approximate NNLO results from \cite{Kidonakis:2010dk} in
Figure~\ref{ptD0tevplot}, using scale variation around the central value 
$\mu_R=\mu_F=m_t=170$ GeV, and MSTW2008 NNLO PDFs. 
The agreement between theory \cite{Kidonakis:2010dk} and experiment is 
very good.
In Figure~\ref{fig:p_t-data} we compare the D0 measurements 
with the NLO and NLO+NNLL calculations from
\cite{Ahrens:2011mw}.  MSTW2008 PDFs are used,
and the theory errors are estimated with scale variations about the 
default values $\mu_R=\mu_F=2m_t$, with $m_t=170$~GeV.  
The NLO+NNLL predictions from \cite{Ahrens:2011mw} are also in good 
agreement with the measurements, although the optimal scale-setting 
procedure for the resummed calculation at higher values of $p_T$ 
where the distribution is small was left as an open point in that work.

\begin{figure}
\begin{center}
\includegraphics[width=10cm]{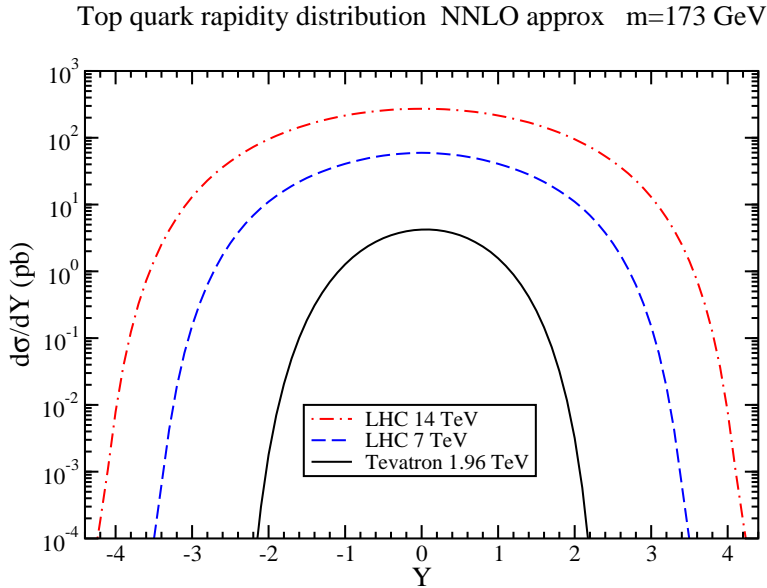}
\caption{The top quark rapidity distribution at the LHC and the Tevatron
from \cite{Kidonakis:2011zn}.}
\label{ylogplot}
\end{center}
\end{figure}

Finally, the NNLO approximate top-quark rapidity distribution at LHC
and Tevatron energies from \cite{Kidonakis:2011zn} is shown in
Fig. \ref{ylogplot}, using MSTW2008 NNLO PDFs with
$\mu_F=\mu_R=m_t=173$ GeV. The rapidity distribution is important for 
the top quark forward-backward asymmetry which we discuss in the next section.

Many more plots showing theoretical
results for the $p_T$ and rapidity distributions can be found in
\cite{Kidonakis:2010dk,Ahrens:2011mw,Kidonakis:2011zn}.

\section{Forward-backward  asymmetry}
\label{sec:FBasy}
Closely related to differential distributions is the top-pair 
forward-backward (FB) asymmetry. Its basic definition is     
\begin{align}
\label{eq:AFBdef}
 A^{i}_{\text{FB}} &= \frac{N(y^i_t>0)-N(y^i_t<0)}{N(y^i_t>0)+N(y^i_t<0)}=
\frac{\sigma(y_t>0)-\sigma(y_t<0)}{\sigma(y_t>0)+\sigma(y_t<0)}\equiv 
\frac{\sigma^i_A}{\sigma^i_S} \, ,
\end{align}
where $N$ is the number of events and $y^i_{t}$ is the top-quark rapidity in
Lorentz frame $i$.  The definition in terms of the number of events is
convenient for experimental measurements, whereas in theory calculations one
obtains the forward-backward asymmetric ($\sigma_A$) and symmetric
($\sigma_S$) cross sections by integrating rapidity distributions over the
appropriate values.  Since each of these are obtained as a series in
$\alpha_s$, one must further decide whether to re-expand the ratio of
asymmetric to symmetric cross sections in the computation of $A_{\rm FB}$, a
point we will come back to in a moment. The basic definition above can also be
modified to include cuts on kinematic variables, for instance by imposing
that $M_{t\bar t}$ or $|y_t|$ is greater or less than a certain value.

At a $p\bar p$ collider such as the Tevatron, charge conjugation invariance in
QCD implies that $N(y^i_{\bar t}>0)=N(y^i_{ t}<0)$, so the FB asymmetry is
equivalent to the charge asymmetry.  At a $pp$ collider such as the LHC,
rapidity distributions are symmetric and the FB asymmetry vanishes.  On the
other hand, the production rates for top and antitop quarks at a
given rapidity are in general different, so partially integrated charge
asymmetries at the LHC do not necessarily vanish.  Moreover, one can define
non-vanishing total asymmetries with respect to variables such as $|\eta_t| -
|\eta_{\bar t}|$, with $\eta$ the pseudorapidity, or $|y_t|-|y_{\bar
t}|$. Various definitions and results for (partially integrated) charge asymmetries
at the LHC can be found elsewhere in this volume \cite{Kamenik:2011wt} and
will not be covered here.  From now on, we will focus solely on the FB asymmetry at
the Tevatron.

Experimental measurements of the FB asymmetry at the Tevatron have been made
in the $t\bar t$ and $p\bar p$ (laboratory) frame.\footnote{Experiment
actually measures the asymmetry with respect to the rapidity difference
$y_t-y_{\bar t}$, but this is equivalent to (\ref{eq:AFBdef}) in the $t\bar t$
frame so we do not distinguish this as a separate observable.}  The
measurements of the total asymmetry obtained by the CDF collaboration using
5.3~fb$^{-1}$ of data are \cite{Aaltonen:2011kc}
$A^{p\bar{p}}_{\text{FB}} = (15.0 \pm 5.5)\% \quad (\text{$p\bar{p}$ frame})$
and
$A^{t\bar{t}}_{\text{FB}} = (15.8 \pm 7.5)\% \quad (\text{$t\bar{t}$ frame})$.
The quoted uncertainties are derived from a combination of statistical and
systematic errors. The measurements are roughly 2$\sigma$ (1$\sigma$) higher
than theoretical results in the laboratory ($t\bar t)$ frame reviewed below.
Even more interesting is the measurement as a function of the top-pair
invariant mass.  After grouping the events in two bins corresponding to
$M_{t\bar{t}} \le 450$~GeV and $M_{t\bar{t}} \ge 450$~GeV,  the
asymmetry in the latter bin was measured to be 
$A^{t\bar{t}}_{\text{FB}} \left(M_{t\bar{t}} \ge 450 \, \mathrm{GeV} \right) 
  = (47.5 \pm 11.4)\%$
which is about 3$\sigma$ higher than the theory predictions in the SM. 
 Attempts to explain the discrepancies of the CDF results with the SM
predictions are reviewed in \cite{Kamenik:2011wt}.  Here, we present the most
up-to-date results for the FB asymmetry within the SM itself, explaining what
is known in fixed-order perturbation theory and within different frameworks
for soft-gluon resummation.

In the SM the FB asymmetry is due mainly to QCD effects.  In
fixed-order perturbation theory the asymmetric cross section first
appears at ${\cal O}(\alpha_s^3)$, in other words at NLO compared to
the symmetric cross section.  The contributions can be traced to
certain types of diagrams in the $q\bar q$ channel and $qg$ channels,
explicitly identified and calculated in \cite{Kuhn:1998jr,
  Kuhn:1998kw}. The $gg$ channel does not contribute to the asymmetric
cross section at any order in perturbation theory, because it involves
a symmetric initial state.  If the ratio $\sigma_A/\sigma_S$ in
(\ref{eq:AFBdef}) is consistently expanded in $\alpha_s$, as we will
assume to be the case unless otherwise indicated, then the
$A_{\rm FB}\sim \alpha_s$ at the first non-vanishing order.  We will
label this leading contribution as NLO, with reference to the order at
which the differential cross section is needed. The total FB asymmetry
at NLO is small in the SM, about $5\%$ in the $p\bar p$ frame and $7\%$
in the $t\bar t$ frame--exact numbers will be given below.  The $q\bar
q$ channel gives the dominant contribution, while that from the $qg$
channel is much smaller numerically.

Given the potential deviations between the measurements and the SM, it
is especially important to estimate the effects of higher-order QCD
corrections on the FB asymmetry. This can be done using the methods of
soft-gluon resummation. We show results for the total FB asymmetry in
the $t\bar t$ and $p\bar p$ frames at NLO and with soft-gluon
resummation included in Table~\ref{tab:fbasym}.  We use MSTW2008 PDFs,
with $m_t=173.1$~GeV for the calculations at NLO+NNLL and
$m_t=173$~GeV at approximate NNLO in the $p\bar p$ frame (the
numerical difference from the mismatch in the top-quark mass is
negligible). The central values refer to the choice $\mu_R=\mu_F=m_t$,
and perturbative errors in fixed order are estimated with correlated
variations by factors of two; in the resummed result also
the matching scales $\mu_h$ and $\mu_s$ are also varied as described
in \cite{Ahrens:2011uf}. In all cases the theory error is much smaller
than the experimental one quoted above, and the main qualitative
finding is that resummation has only a rather mild effect on the FB
asymmetry.  This is true whether one uses NLO+NNLL, or the approximate
NNLO calculations of \cite{Ahrens:2010zv, Ahrens:2011mw,
  Kidonakis:2011zn}.\footnote{In comparing results it is important to
  note that \cite{Kidonakis:2011zn} does {\it not} expand the FB
  asymmetry in $\alpha_s$, evaluating instead the denominator of
  (\ref{eq:AFBdef}) numerically at NNLO with no further expansion. }
In fact, the same conclusions hold for the invariant-mass dependent
CDF analysis, where the asymmetry is studied in a bin below and above
a cut at $M_{t\bar t}=450$~GeV.  The studies in \cite{Ahrens:2011uf}
showed that such a binning roughly divides the total asymmetric cross
section in half, and that the effect of resummation in each bin is
roughly the same as for the results shown in the table here--in other
words, it is a mild effect.  Similar conclusions can be drawn from the
NLO+NLL study of \cite{Almeida:2008ug}.

\begin{table}
  \centering
  \begin{tabular}{|l|c|c|}
    \hline
    & $A^{t\bar t}_{\text{FB}}$ [\%]
    & $A^{p\bar p}_{\text{FB}}$ [\%]
    \\ \hline
    NLO \cite{Ahrens:2011uf} 
    &  7.32{\footnotesize $^{+0.69+0.18}_{-0.59-0.19}$}
    &   4.81{\footnotesize $^{+0.45+0.13}_{-0.39-0.13}$}
    \\ \hline 
    NLO+NNLL \cite{Ahrens:2011uf} 
    & 7.24{\footnotesize $^{+1.04+0.20}_{-0.67-0.27}$}
    &  4.88{\footnotesize $^{+0.20+0.17}_{-0.23-0.18}$}
    \\ \hline  
    NNLO$_{\rm approx.}$ \cite{Kidonakis:2011zn} 
    & ---
    & 5.2{\footnotesize $^{+0.0}_{-0.6}$}
    \\ \hline  
  \end{tabular}
  \caption{\label{tab:fbasym} The FB asymmetry in the $t\bar{t}$ and 
    $p\bar p$ rest frames. The first error 
    refers to perturbative uncertainties estimated through scale
    variations, the second (where applicable) to PDF uncertainties.}
\end{table}

The moderate size of the higher-order QCD corrections as estimated by
soft gluon resummation makes it important to consider other sources of
uncertainty in the SM calculation.  First, there are of course subleading
terms in the soft limit to which soft-gluon resummation has no access.  The
magnitude of these terms is estimated through scale variation but will be known for sure only once the fixed-order calculation is
done at the next order.  Second, electroweak corrections (EW) can also be
important for the FB asymmetry.  These were considered already in
\cite{Kuhn:1998kw}, where they were estimated to increase the asymmetry
roughly by a factor of $1.1$. Refinements related to additional contributions
were made in \cite{Bernreuther:2010ny}, where a slightly smaller contribution
was found, but very recently the calculations of \cite{Hollik:2011ps} quote an
increase of about $1.2$ due to EW corrections.  The corrections for binned 
analyses in $M_{t \bar t}$ and pair rapidity difference were found to be
roughly the same in the bins studied by experiment. These corrections are
comparable or larger than the QCD uncertainties and thus cannot be
ignored--however, they do little to change the 3$\sigma$ discrepancy between
the SM and the 
CDF measurement in the high invariant-mass bin.

\section{Single top production}
\label{sec:singletop}

The observation of single top quark production at the Tevatron
\cite{Abazov:2009ii,Abazov:2009pa,Abazov:2009nu,Abazov:2011rz,Aaltonen:2009jj,Aaltonen:2010jr,Group:2009qk} and at the LHC
\cite{Chatrchyan:2011vp, ATLAStch} has
increased the need for reliable theoretical calculations of the cross 
sections for the corresponding production processes.
The single top cross section is less than half of that for $t{\bar t}$ 
production but the backgrounds are considerable thus making the observation 
of single tops quite challenging.

\begin{figure}
\begin{center}
\includegraphics[width=3cm]{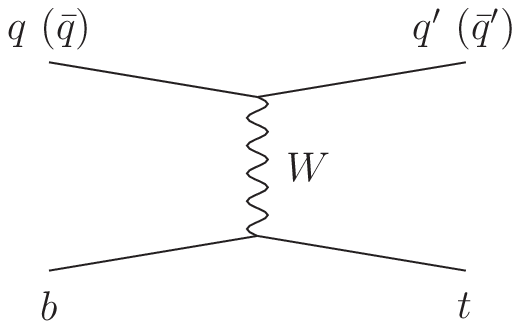} \hspace{8mm}
\includegraphics[width=3cm]{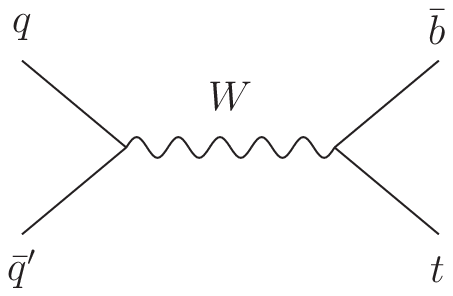} \hspace{8mm}
\includegraphics[width=7cm]{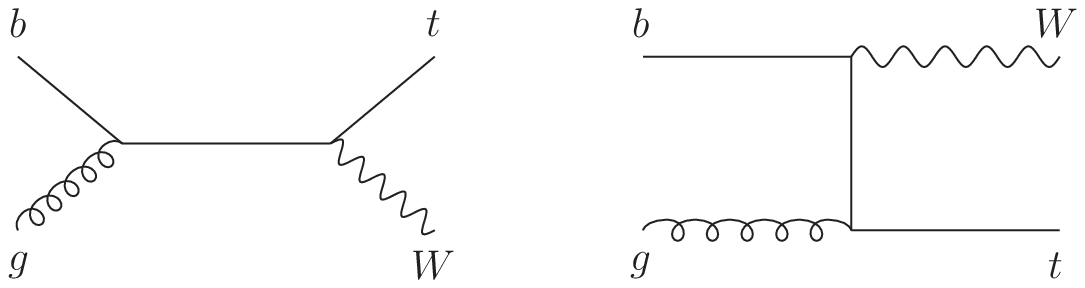}
\caption{\label{singletoplo} Leading-order $t$-channel (left), $s$-channel, and $tW$ (right two)  diagrams for single top quark production.}

\end{center}
\end{figure}

Single top quarks can be produced through three distinct
partonic processes. One of them is the $t$-channel process that proceeds via
the exchange of a space-like $W$ boson, a second is the $s$-channel
process that proceeds via the exchange of a time-like $W$ boson,
and a third is associated $tW$ production.
The leading-order diagrams are shown in Fig. \ref{singletoplo}.

\begin{table}[h]
\begin{center}
  \begin{tabular}{c|c|c}
LHC 7 TeV & $t$ &  ${\bar t}$ \\ \hline
$t$-channel  \cite{Kidonakis:2011wy} & 
  $41.7^{+1.6}_{-0.2}\pm 0.8$ &
  $22.5 \pm 0.5 {}^{+0.7}_{-0.9}$ 
\\
$s$-channel \cite{Kidonakis:2010tc} & 
  $3.17 \pm 0.06 {}^{+0.13}_{-0.10}$ &
  $1.42 \pm 0.01 {}^{+0.06}_{-0.07}$ 
\\ 
$tW$ \cite{Kidonakis:2010ux} &
 $7.8 \pm 0.2 {}^{+0.5}_{-0.6}$ &
 $7.8 \pm 0.2 {}^{+0.5}_{-0.6}$
\end{tabular}
\end{center}
\vspace{-2mm}
\caption{\label{tab:singletop} Results for single-top and single-antitop 
approximate NNLO cross sections at the LHC with $m_t=173$ GeV in pb. The first 
uncertainty is from scale variation and the second is the PDF error 
using the MSTW2008 NNLO PDF sets at 90\% CL \cite{Martin:2009iq}.  }
\end{table}

\subsection{$t$-channel production}

The $t$-channel partonic processes are $qb \rightarrow q' t$
and ${\bar q} b \rightarrow {\bar q}' t$.
At both the LHC and the Tevatron the $t$-channel is numerically dominant.
Calculations of NLO corrections for $t$-channel 
production at the differential level have been known for some time 
\cite{Harris:2002md}
(see also further recent NLO studies in 
\cite{Campbell:2009ss,Campbell:2009gj,Falgari:2010sf,Schwienhorst:2010je,Falgari:2011qa}).

Theoretical calculations for single top quark production
beyond NLO that include higher-order corrections
from soft-gluon resummation
appeared in \cite{Kidonakis:2006bu,Kidonakis:2007ej} at NLL and in 
\cite{Kidonakis:2011wy} at NNLL. 
The NNLL theoretical expressions in \cite{Kidonakis:2011wy} were used 
to derive approximate NNLO cross sections for $t$-channel
single top or single antitop production at the Tevatron and the LHC.

\begin{figure}
\begin{center}
\includegraphics[width=10cm]{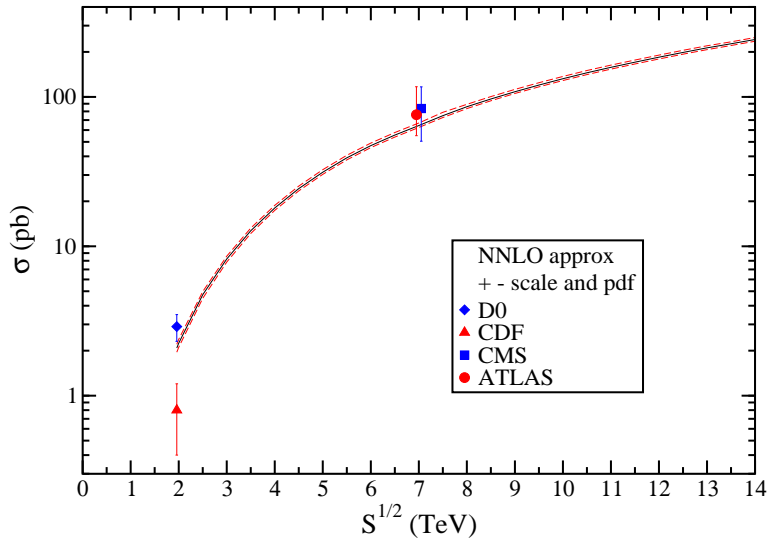}
\caption{The single top plus single antitop $t$-channel cross section
versus collider energy.}
\label{tchS2011plot}
\end{center}
\end{figure}

In Figure \ref{tchS2011plot} the $t$-channel approximate NNLO cross section
from NNLL resummation is plotted versus collider energy, together with 
recent measurements from the Tevatron \cite{Abazov:2011rz,Aaltonen:2010jr} 
and LHC experiments \cite{Chatrchyan:2011vp,ATLAStch}.
For a top quark mass of 173 GeV the $t$-channel single top quark 
NNLO approximate cross section at the Tevatron is
\beq
\sigma_{\rm t-ch}^{\rm top}(m_t=173\, {\rm GeV}, \, \sqrt{s}=1.96\, {\rm TeV})
=1.04^{+0.00}_{-0.02} \pm 0.06  \; {\rm pb}
\eeq
where the first uncertainty is from scale variation between $m_t/2$ and $2m_t$
and the second is the PDF uncertainty, calculated using the 
MSTW2008 NNLO PDF sets \cite{Martin:2009iq} at 90\% C.L.
We note that the results for
single antitop production at the Tevatron are identical to those for 
single top.

For $t$-channel production at the LHC we note 
that the single top cross section is different from that for single 
antitop production.
For $m_t=173$ GeV we list the results for LHC 7 TeV energy 
in Table \ref{tab:singletop}.
 
\subsection{$s$-channel production}

The lowest-order processes in the $s$-channel are of the form
$q{\bar q}' \rightarrow {\bar b} t$, which include
the dominant process $u {\bar d} \rightarrow {\bar b} t$
as well as processes involving the charm quark and Cabibbo-suppressed
contributions.
The QCD corrections for $s$-channel production at next-to-leading order (NLO)
are known at the differential level \cite{Harris:2002md} and they 
substantially increase the cross 
section and stabilize the dependence on the factorization scale.

Approximate NNLO calculations from NLL soft-gluon 
resummation appeared in \cite{Kidonakis:2006bu,Kidonakis:2007ej} 
and from NNLL resummation in \cite{Kidonakis:2010tc}.
The soft-gluon corrections dominate the cross section and the NLO expansion 
of the resummed cross section approximates very well the complete NLO result
for both Tevatron and LHC energies. 
Results based on SCET have appeared in \cite{Zhu:2010mr}. 
In addition to the difference in formalism,  
Refs. \cite{Kidonakis:2010tc} and \cite{Zhu:2010mr} differ in their 
definitions of the $s_4$ variable in the resummation; we refer 
the reader to the papers for a more detailed discussion.  
Here we just note that contrary to the study in
\cite{Kidonakis:2010tc} the NLO threshold expansion 
in the approach of  \cite{Zhu:2010mr} was found  to poorly approximate
the full corrections at LHC energies.

For a top quark mass of 173 GeV, the $s$-channel single top cross section
and its associated uncertainties at the Tevatron are \cite{Kidonakis:2010tc}
\beq
\sigma_{\rm s-ch}^{\rm top}(m_t=173\, {\rm GeV}, \, \sqrt{s}=1.96\, {\rm TeV})
=0.523^{+0.001}_{-0.005}{}^{+ 0.030}_{-0.028} \; {\rm pb}
\eeq
where the first uncertainty is from scale variation and the second is
the PDF uncertainty at 90\% C.L.
The results for single antitop production at the Tevatron are identical.

The results for single top and single-antitop $s$-channel production at the 
LHC at 7 TeV with $m_t=173$ GeV are shown in Table \ref{tab:singletop}.

\subsection{$tW$ production}

The associated production of a top quark with a
$W$ boson, $bg \rightarrow tW^-$, is sensitive to new physics
and allows a direct measurement of the $V_{tb}$ CKM matrix element.
This process is practically negligible at the Tevatron but it has the second
highest cross section among single top processes at the LHC.
A similar process in physics beyond the Standard Model is associated 
production of a top quark with a charged Higgs boson, $bg \rightarrow tH^-$. 
The next-to-leading order (NLO) corrections to $bg \rightarrow tW^-$ were 
calculated in \cite{Zhu:2002uj}.

NNLO soft-gluon corrections from NLL resummation were calculated in
\cite{Kidonakis:2006bu,Kidonakis:2007ej} and from NNLL resummation in 
\cite{Kidonakis:2010ux}.
The NLO expansion of the
resummed cross section approximates well the complete NLO result
for both Tevatron and LHC energies. 

The results for $tW^-$ and ${\bar t} W^+$ at the 
LHC at 7 TeV with $m_t=173$ GeV are shown in Table \ref{tab:singletop}.
The NNLO approximate corrections increase the NLO cross
section by $\sim 8$\%. 
We note that the cross section for
${\bar b} g \rightarrow {\bar t} W^+$ is identical
at both Tevatron and LHC to that for $bg \rightarrow tW^-$.

\section{Conclusions}
\label{sec:conclusions}

We have reviewed QCD calculations for inclusive top-quark production at
Tevatron and LHC energies.  The total and differential cross sections have
been known at NLO in QCD for many years, and more recently different
implementations of soft-gluon resummation at NNLL order have also become
available and applied to many observables.  Compared to NLO calculations,
phenomenological predictions based on resummation invariably show smaller
dependence on factorization and renormalization scales, and produce moderate
changes with respect to central values for a default scale choice. However,
especially for the total pair-production cross section at the Tevatron,
different forms of resummation can yield results which show only nominal
agreement with one another within the quoted uncertainties (see
Table~\ref{tab:cspole}).  We explained in detail the reasons for such
differences at the level of approximate NNLO results, but at present the best
way of implementing resummation for the total cross section is still subject
to debate, as is the most reliable way of estimating the uncertainties.
Ultimately, only a complete NNLO calculation will resolve this point.

\section*{Acknowledgements}

The work of N.K. was supported by the National Science Foundation under 
Grant No. PHY 0855421.

\end{document}